\documentclass[8pt,osajnl2,preprint,superscriptaddress]{revtex4-1}

\usepackage[utf8]{inputenc}
\usepackage[T1]{fontenc}
\usepackage{amsmath}
\usepackage{physics}
\usepackage{graphicx}
\usepackage{stmaryrd}
\usepackage{float}
\usepackage{xcolor}
\usepackage{soul}

\usepackage{pdfpages}
\usepackage{etoolbox} 

\makeatletter
\patchcmd{\@outputpage@head}{\@ifx{\LS@rot\@undefined}{}{\LS@rot}}{}{}{}
\makeatother

\usepackage[bottom]{footmisc}

\usepackage[labelsep=endash]{caption}
\newcommand\svdots{{\vphantom{\int^0}\smash[t]{\vdots}}}
\newcommand\sddots{{\vphantom{\int^0}\smash[t]{\ddots}}}
\newcommand\shdots{{\vphantom{\int^0}\smash[t]{\hdots}}}

\usepackage{letltxmacro}
\LetLtxMacro\orgvdots\vdots
\LetLtxMacro\orgddots\ddots

\makeatletter
\DeclareRobustCommand\vdots{%
  \mathpalette\@vdots{}%
}
\newcommand*{\@vdots}[2]{%
  \sbox0{$#1\cdotp\cdotp\cdotp\m@th$}%
  \sbox2{$#1.\m@th$}%
  \vbox{%
    \dimen@=\wd0 %
    \advance\dimen@ -3\ht2 %
    \kern.5\dimen@
    \dimen@=\wd2 %
    \advance\dimen@ -\ht2 %
    \dimen2=\wd0 %
    \advance\dimen2 -\dimen@
    \vbox to \dimen2{%
      \offinterlineskip
      \copy2 \vfill\copy2 \vfill\copy2 %
    }%
  }%
}
\DeclareRobustCommand\hdots{%
  \mathinner{%
    \mathpalette\@hdots{}%
    \mkern\thinmuskip
  }%
}
\newcommand*{\@hdots}[2]{%
  \sbox0{$#1\cdotp\cdotp\cdotp\m@th$}%
  \sbox2{$#1.\m@th$}%
  \vbox{%
    \dimen@=\wd0 %
    \advance\dimen@ -3\ht2 %
    \kern.5\dimen@
    \dimen@=\wd2 %
    \advance\dimen@ -\ht2 %
    \dimen2=\wd0 %
    \advance\dimen2 -\dimen@
    \vbox to \dimen2{%
      \offinterlineskip
      \vfill
      \hbox{$#1\mathpunct{.}\mathpunct{.}\mathpunct{.}\m@th$}%
      \vfill
    }%
  }%
}
\DeclareRobustCommand\ddots{%
  \mathinner{%
    \mathpalette\@ddots{}%
    \mkern\thinmuskip
  }%
}
\newcommand*{\@ddots}[2]{%
  \sbox0{$#1\cdotp\cdotp\cdotp\m@th$}%
  \sbox2{$#1.\m@th$}%
  \vbox{%
    \dimen@=\wd0 %
    \advance\dimen@ -3\ht2 %
    \kern.5\dimen@
    \dimen@=\wd2 %
    \advance\dimen@ -\ht2 %
    \dimen2=\wd0 %
    \advance\dimen2 -\dimen@
    \vbox to \dimen2{%
      \offinterlineskip
      \hbox{$#1\mathpunct{.}\m@th$}%
      \vfill
      \hbox{$#1\mathpunct{\kern\wd2}\mathpunct{.}\m@th$}%
      \vfill
      \hbox{$#1\mathpunct{\kern\wd2}\mathpunct{\kern\wd2}\mathpunct{.}\m@th$}%
    }%
  }%
}
\makeatother

\begin{document}
  
  %%%%%%%%%%%%%%%%%% title page information %%%%%%%%%%%%%%%%%%
  \title{Turning Optical Complex Media into Universal  Reconfigurable Linear Operators by Wavefront Shaping}
  
  \author{Maxime W. Matthès}
  \affiliation{Institut Langevin, CNRS UMR 7587, ESPCI Paris, PSL Research University, 1 rue Jussieu, 75005 Paris, France}
  \author{Philipp del Hougne}
  \affiliation{Institut Langevin, CNRS UMR 7587, ESPCI Paris, PSL Research University, 1 rue Jussieu, 75005 Paris, France}
  \author{Julien de Rosny}
  \affiliation{Institut Langevin, CNRS UMR 7587, ESPCI Paris, PSL Research University, 1 rue Jussieu, 75005 Paris, France}
  \author{Geoffroy Lerosey}
  \affiliation{Greenerwave, ESPCI Paris Incubator PC’up, 6 rue Jean Calvin, 75005 Paris, France}
  \author{Sébastien M. Popoff}
  \email{sebastien.popoff@espci.psl.eu}
  \affiliation{Institut Langevin, CNRS UMR 7587, ESPCI Paris, PSL Research University, 1 rue Jussieu, 75005 Paris, France}

\begin{abstract}

  Performing linear operations using optical devices is a crucial building block in many fields ranging from telecommunication to optical analogue computation and machine learning. For many of these applications, key requirements are robustness to fabrication inaccuracies and reconfigurability. Current designs of custom-tailored photonic devices or coherent photonic circuits only partially satisfy these needs.
  Here, we propose a way to perform linear operations by using complex optical media such as multimode fibers or thin scattering layers as a computational platform driven by wavefront shaping. Given a large random transmission matrix (TM) representing light propagation in such a medium, we can extract a desired smaller linear operator by finding suitable input and output projectors. 
  We discuss fundamental upper bounds on the size of the linear transformations our approach can achieve and provide an experimental demonstration. For the latter, first we retrieve the complex medium's TM with a non-interferometric phase retrieval method. Then, we take advantage of the large number of degrees of freedom to find input wavefronts using a Spatial Light Modulator (SLM) that 
  cause the system, composed of the SLM and the complex medium, to act as a desired complex-valued linear operator on the optical field. We experimentally build several $16\times16$ complex-valued operators, and are able to switch from one to another at will. Our technique offers the prospect of reconfigurable, robust and easy-to-fabricate linear optical analogue computation units.

\end{abstract}  

\maketitle  

\section*{Introduction}

The ability to perform linear operations on light with optical devices is a fundamental ingredient in many areas of optics and photonics, including signal processing and spatial multiplexing in optical communication, as well as optical analogue computation. For emerging applications of the latter to optical artificial neural networks, the ability to reconfigure these linear photonic processing units is crucial. Moreover, ease of fabrication and robustness to manufacturing inaccuracies are highly sought-after features.
Traditionally, photonic devices, similarly to electronic devices, are designed to perform one given operation \cite{borel2007imprinted,jensen2011topology,frellsen2016topology,silva2014performing,piggott2015inverse,shen2015integrated,piggott2017fabrication}.
The conformation of the device is directly linked to the function it is designed for.
As a consequence, fabrication imperfections and changes of environmental conditions negatively impact its functioning, limiting the range of operation of the device. 
Furthermore, such inverse-design approaches inherently prohibit reconfigurability. 

Programmable coherent photonic circuits have the potential to offer reconfigurability. Initially conceived in free-space optics by leveraging beam-splitters \cite{reck94}, recent advances in silicon photonics also enabled the implementation of the concept in integrated designs \cite{miller2013self-configuring,carolan2015universal,ribeiro2016demonstration,annoni_unscrambling_2017,shen2017deep,miller2017silicon,optica18}. 
Despite several promising first experimental demonstrations of this approach, the scaling of the amount of required phase-shifters in these architectures with the square of the size of the desired transformation raises questions about the overall scalability of the approach. 
Another approach using multi-plane modulation was proposed~\cite{morizur2010programmable} and was successfully utilized for few mode manipulations~\cite{labroille2014efficient}.
While small transformations may suffice for optical communications, the increasing complexity of the physical system may hinder the implementation of the high rank transformations needed in optical analogue computation.

In parallel to the aforementioned developments, since Vellekoop and Mosk's landmark paper in 2007 \cite{vellekoop_focusing_2007} the field of wavefront shaping in complex media emerged~\cite{mosk_controlling_2012,rottergigan}.
A complex medium may be defined as a system that mixes the spatial and/or temporal degrees of freedom, resulting in the complete scrambling of an incident wavefront \cite{goodman2007speckle}. Examples include chaotic cavities, disordered waveguides or random scattering systems such as paint layers or biological tissues \cite{vellekoop_focusing_2007,mosk_controlling_2012,vcivzmar2011shaping,kim2015transmission,sarma2015control,doya2001light,doya2002speckle,draeger1997one,dupre2015wave,del2016intensity,bittner2018suppressing}.
The seemingly random effect of the disorder on the wavefront is  deterministic; hence such a system can be fully represented by a linear transmission matrix (TM) \cite{popoff_measuring_2010}.
Due to the large number of modes, those matrices have large dimensions. Moreover, the lack of symmetry of the system contributes to the TM's high rank.
By injecting light into a well-chosen combination of input modes using a spatial light modulator (SLM), an optical scattering medium can be utilized to perform a wide variety of functions. Initial efforts sought to overcome the scrambling of the wavefront in space and/or time through wavefront shaping, e.g. to focus behind or transmit information through a complex medium \cite{vellekoop_focusing_2007,popoff_measuring_2010,popoff_image_2010,dupre2015wave,del2016intensity,aulbach2011control,katz2011focusing,mounaix2016spatiotemporal,del2016spatiotemporal}. Strikingly, it was possible to beat the Rayleigh limit on focusing \cite{vellekoop_exploiting_2010,choi2011overcoming,park_subwavelength_2013} and the Nyquist-Shannon sampling theorem \cite{liutkus_imaging_2013}. 

Yet controlling wave propagation in complex media to perform linear transformations has received %surprisingly 
little attention despite its potential. 
The literature contains reports on implementations of two-port beam-splitters \cite{huisman2014programming,huisman2015programmable}, with applications to control quantum interferences in mind \cite{defienne2016two-photon,wolterink2016programmable}, as well as on spatial mode sorting \cite{fickler2017custom-tailored}. 
In particular, the possibility of implementing large reconfigurable linear transformations as required for optical neural networks has remained unexplored to date. 
Building on recent work that demonstrated wave-based analogue computation in a chaotic microwave cavity \cite{dH_WBAC}, we explore here the possibility of performing large complex-valued linear operations in optical complex media. We unlock the potential of linear optical signal processing with shaped optical waves in complex media by formulating fundamental upper bounds in terms of the  operation size that can be implemented in a realistic optical system. 
Moreover, we experimentally demonstrate the possibility to take advantage of the large number of degrees of freedom of a disordered optical system (multimode fiber or glass diffuser) 
of which we acquire the TM, to physically perform different desired linear transformations. 
We do not rely on interferometric phase measurements but leverage phase retrieval techniques to show that our scheme can be implemented with minimal hardware requirements. Moreover, the one-off calibration of the medium's TM allows us to reconfigure the implemented linear transformation without any further experimental measurement.

\section*{Results}

\textbf{Principle.} Light propagation through a linear disordered system at a given frequency is fully described by its transmission matrix \textbf{H} that links the output state of light $\ket{\psi_{out}}$ to the input one $\ket{\psi_{in}}$:

\begin{equation}
\ket{\psi_{out}} = \mathbf{H}\ket{\psi_{in}}.
\label{eq:TM}
\end{equation}

\noindent Usually only a small subset of the input and output modes are controlled and measured. Then, the transmission matrix has the statistical properties of an ideal random Gaussian matrix~\cite{goetschy2013filtering,Popoff_2014}.
If one controls $N$ input modes and measures $M$ output modes, $\mathbf{H}$ is a $M \times N$ matrix and $\ket{\psi_{in}}$ (resp $\ket{\psi_{out}}$) is represented by a vector of size $N$ (resp. $M$).
Suppose we seek to create a system performing a linear operation represented by the matrix $\mathbf{G}$ of size $m \times n$ with $m < M$ and $n < N$. Having measured the complex medium's transmission matrix $\mathbf{H}$, we can identify adequate input and output projections, represented by $N \times n$ and $M \times m$ input and output matrices $\mathbf{P_{in}}$ and $\mathbf{P_{out}}$, that satisfy:
\begin{equation}
\mathbf{G} = \mathbf{P_{out}^{T}}\mathbf{H}\mathbf{P_{in}}.
\label{eq:A}
\end{equation}

\begin{figure}[ht]
	\centering
	\includegraphics[width= 0.9\textwidth ]{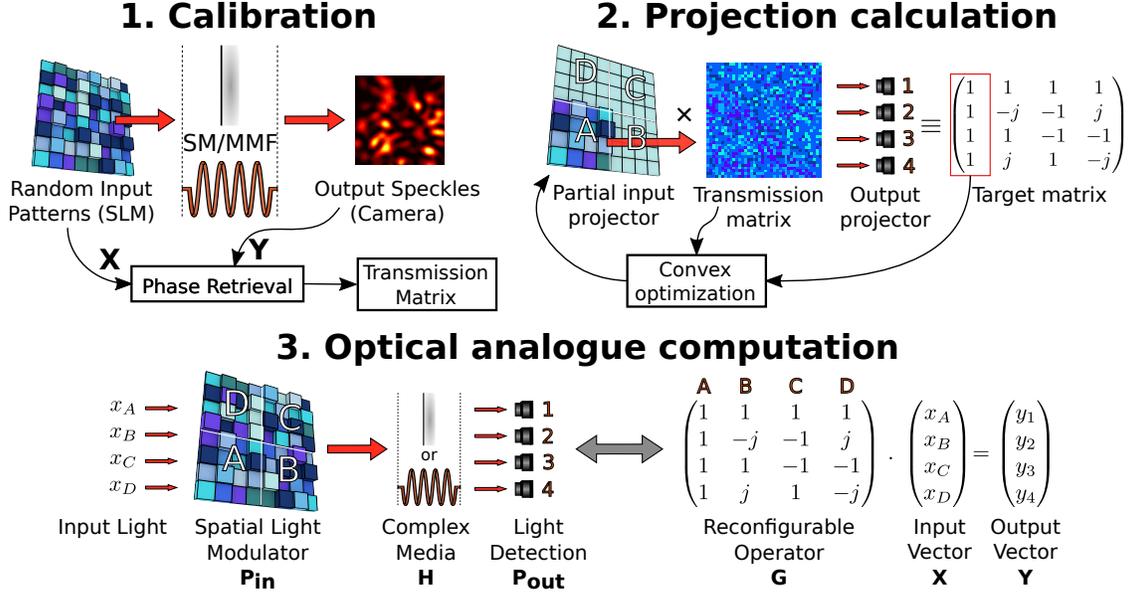}
	\caption{Overview of the experimental procedure for performing a DFT operation of size $4$ ($n=m=4$ and $\mathbf{G} = \mathrm{DFT_4}$). 
		\textbf{Step~1: Calibration of the system.}
		Acquisition of the complex TM by measuring a set of input patterns and output intensity speckles and using a phase retrieval algorithm. 
		\textbf{Step~2: Calculation of the optimal input projection.}
		We illustrate the procedure by showing how to find one subpart of the input mask.
		The computation is done independently for each subpart of the SLM using a convex optimization solver and using the TM and the corresponding column of the target matrix $\mathbf{G}$ as inputs.
		\textbf{Step~3: Analogue computation.} 
		The proposed optical processing unit is composed of a spatial light modulator (SLM) and a complex medium.
		The SLM and the output detection take the role of the projectors $\mathbf{P_{in}}$ and $\mathbf{P_{out}}$, converting the given transmission matrix $\mathbf{H}$ of the complex medium into a desired linear transformation $\mathbf{G}$ (see~equation~\ref{eq:A}). 
	}
	\label{post_optim}
\end{figure}

\noindent Experimentally, we use a spatial light modulator to impose the input projector $\mathbf{P_{in}}$.
We divide the modulator into $n$ groups of $N/n$ pixels on which we control the amplitude and/or the phase of the optical field.
The output projection $\mathbf{P_{out}}$ is realized by measuring $m$ speckle grains on the camera (see Methods for a mathematical expression of these projectors). 
This concept is schematically illustrated in figure~\ref{post_optim}.

The operation is performed on a desired input vector by encoding its components $ \mathbf{X} = (x_1, x_2, ..., x_n)^T $ into as many light beams directed towards the SLM and collecting the output vector $\mathbf{Y}$ on the detection device such that: 
\begin{equation*}
    \mathbf{Y} = \mathbf{G}\mathbf{X} = \mathbf{P_{out}^{T}}\mathbf{H}\mathbf{{P}_{in}}\mathbf{X}.
\end{equation*}

\textbf{All optical operations.}

To illustrate the versatility of our approach, we report two experimental implementations of our scheme using two very different complex media as optical processing units: a multimode fiber and a ground glass diffuser at two different wavelengths (respectively 1550 nm and 632.8 nm).
We demonstrate the universality of the presented scheme by experimentally implementing two linear transformations  very common in computer and physical science, namely the discrete Fourier transform and the Hadamard matrix. 
Their general expressions read:
\begin{equation*}
\mathbf{DFT}_\text{n} =\frac{1}{\sqrt{n}}
\begin{bmatrix}
\omega^{jk}_\text{n}
\end{bmatrix}_{j,k=0..n-1}
\text{ with } \omega_\text{n} = e^{-2\pi i/n},
\end{equation*}

and

\begin{equation*}
\mathbf{Ha}_\text{n+1} = 
\begin{bmatrix}
\text{Ha}_\text{n} & \text{Ha}_\text{n} \\
\text{Ha}_\text{n} & -\text{Ha}_\text{n} 
\end{bmatrix}
\text{ with Ha}_1=
\begin{bmatrix}
1
\end{bmatrix}.
\end{equation*}
with  $n$ the size of the operator.
We quantify the quality of the operation by estimating the operator's fidelity $F_c = Tr(|\widetilde{\mathbf{G}}.\mathbf{G}^ \dagger |^2)*n$, with $\widetilde{\mathbf{G}}$ the response matrix of our physical system after projection and $\mathbf{G}$ the target matrix ($\text{DFT}_\text{n}$ or $\text{Ha}_\text{n}$).
To characterize the implemented operator $\widetilde{\mathbf{G}}$, we sequentially send vectors from an input basis and measure the corresponding outputs fields. 
To measure the complex output field, we add an off-axis interferometric arm to the setup. 
Note that this arm was not part of the procedure of implementing the linear operator, it is solely used to monitor the effective complex-valued operator $\widetilde{\mathbf{G}}$ afterwards~\cite{Cuche_99} (See Supplementary Methods for details about the complex field measurements).

 \begin{figure}[H]
	\centering
	\includegraphics[width= 0.40\textwidth ]{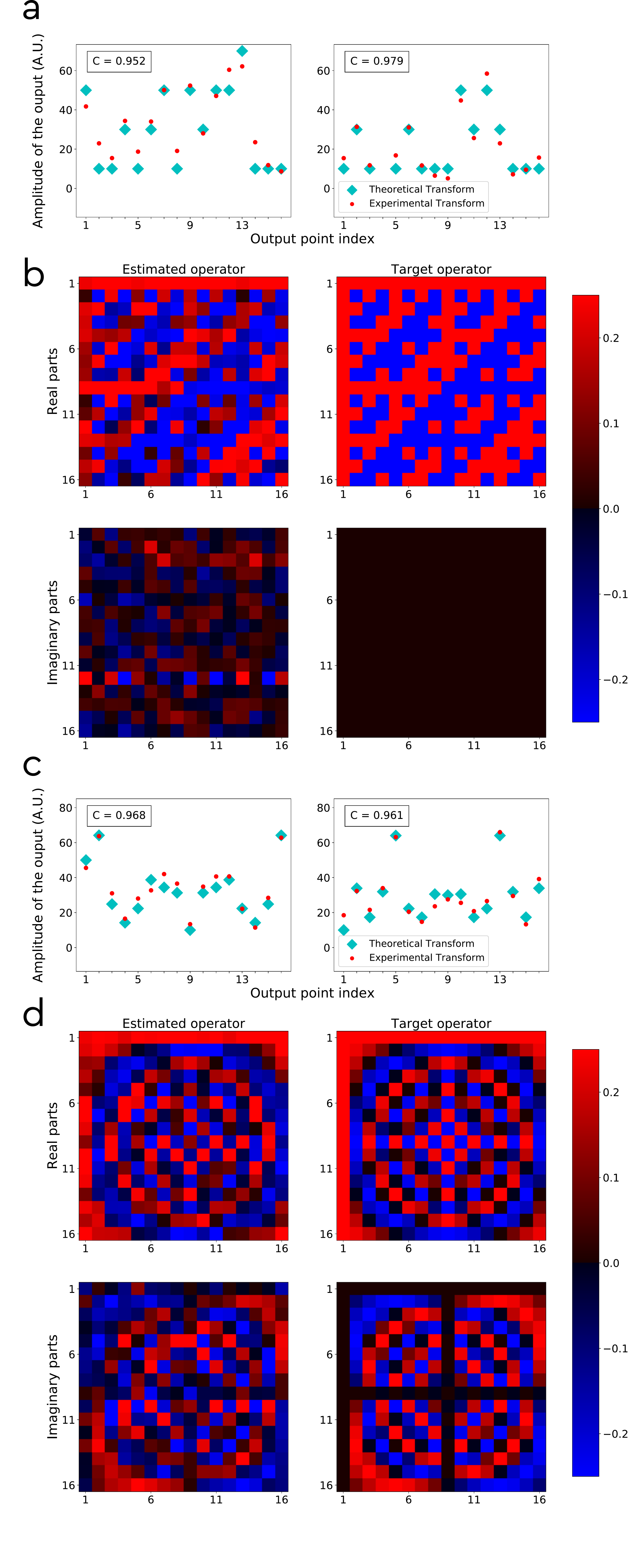}
	\caption{Comparison of experimentally implemented operators and target operators. 
		We use as target matrix $\mathbf{G}$ a $16$ by $16$ Hadamard matrix  (\textbf{a} and \textbf{b}) and a $16$ by $16$ discrete Fourier transform (\textbf{c} and \textbf{d}). 
		Absolute values of experimental (red dots) and theoretical (cyan diamonds) results of the linear transformations for 2 different random input vectors drawn from $\{-1,0,1\}$ for $\mathbf{G} = \text{Ha}_{16}$ (\textbf{a}) and $\mathbf{G}=\text{DFT}_{16}$ (\textbf{c}). 
		The correlation between the two signals is shown in insert. 
		Comparison between experimental and target operators for $\mathbf{G} = \text{Ha}_{16}$ (\textbf{b}) and $\mathbf{G}=\text{DFT}_{16}$ (\textbf{d}).
		The results are obtained without averaging, more results can be found in Supplementary Methods.}
	\label{MMF_data}
\end{figure}

\begin{figure}[ht]
	\centering
	\includegraphics[width= 0.7\textwidth ]{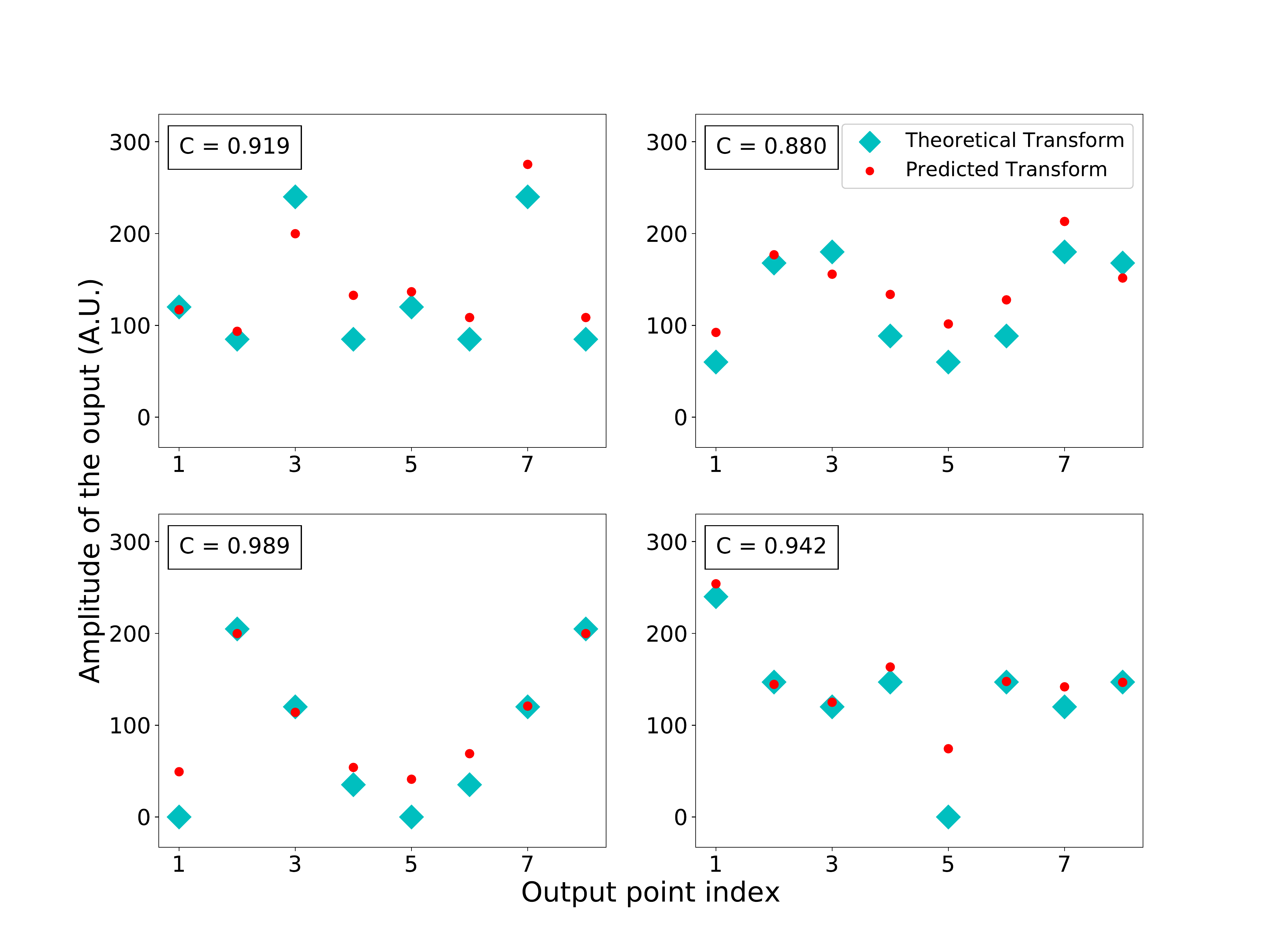}
	\caption{Amplitude of four different random output vectors for operator $\mathbf{G} = \text{DFT}_{8}$ obtained using a ground glass diffuser as a complex medium. 
		\textit{Cyan} diamonds correspond to the theoretical outputs and \textit{red} dots to the experimental data.}
	\label{SM_raw_data}
\end{figure}

We experimentally perform all optical operations according to the presented principle. 
A large range of input signals is prepared and sent to the setup (see the detailed procedure in Supplementary Methods). 
For the sake of simplicity, and without loss of generality, we modulate the input vector $\mathbf{X}$ using the same modulator as the one used for generating the input projector $\mathbf{P_{in}}$  (see the detailed procedure in Supplementary Methods).
We present in figure~\ref{MMF_data} examples of amplitude measurements for optical operations through a multimode fiber, as well as the complex representations of theoretical and measured operators. 
In figure~\ref{SM_raw_data}, we show amplitude measurements for the scattering medium setup. All of these results are obtained without averaging.
We compare these values with the output values given by the true operator. 
The quality of our results is assessed by the Pearson correlation coefficient between the absolute values of the experimental and the ideal output vectors $\mathbf{C}$
and also by the experimental fidelity $\mathbf{F_c}$. 
A summary of the measured values for these quantities for the MMF experiment is presented in table~\ref{tab:res} with and without averaging for operator sizes of $n=m=8$ and $n=m=16$.
A good agreement between the experimental data and the ideal operations outputs is observed for the different sizes tested, even without averaging.
It demonstrates the possibility of performing \textit{one shot} operations through a multimode fiber. 
To further demonstrate the possibility to create any desired operator, we show in Supplementary Methods an experimental implementation of a matrix displaying the name of the author's host institution by encoding information independently in the real and imaginary part of its elements.

\begin{table}
  \centering
  \begin{tabular}{|c|c|c|c|}
    \hline
    Size & Averaging & $\mathbf{C}$ & $\mathbf{F_c}$\\
    \hline
    $n=m=8$ & 1 & $ 0.977 \pm 0.009 $ & $0.973 \pm 0.014$ \\
    \hline
    $n=m=8$ & 10 & $ > 0.99 $ & $0.996$ \\
    \hline
    $n=m=16$ & 1 &  $0.912 \pm 0.027$ & $0.79 \pm 0.027$  \\
    \hline
    $n=m=16$ & 10 &  $> 0.99$ & $0.968$\\
    \hline
  \end{tabular}
  \caption{Summary of the efficiency results for the MMF experiment.} \label{tab:res}
\end{table}

The ground glass diffuser experiment is more prone to errors in the operator reconstruction, we therefore only present results for $n=m=8$. In figure~\ref{SM_raw_data} we show the raw amplitude of 4 different outputs for the $\text{DFT}_8$ operator. We measure $\mathbf{C} = 0.948\pm 0.018$. 
We estimate that the sources of the errors are the lower laser stability and the non-linear response of the CCD camera.
While negatively impacting the results, we obtain qualitatively similar results, demonstrating the capacity of the approach to work with relatively cheap devices, namely a HeNe laser and a standard CCD camera. 
More results are displayed in the Supplementary Methods.

\begin{figure}[ht]
	\includegraphics[width= 0.60\textwidth ]{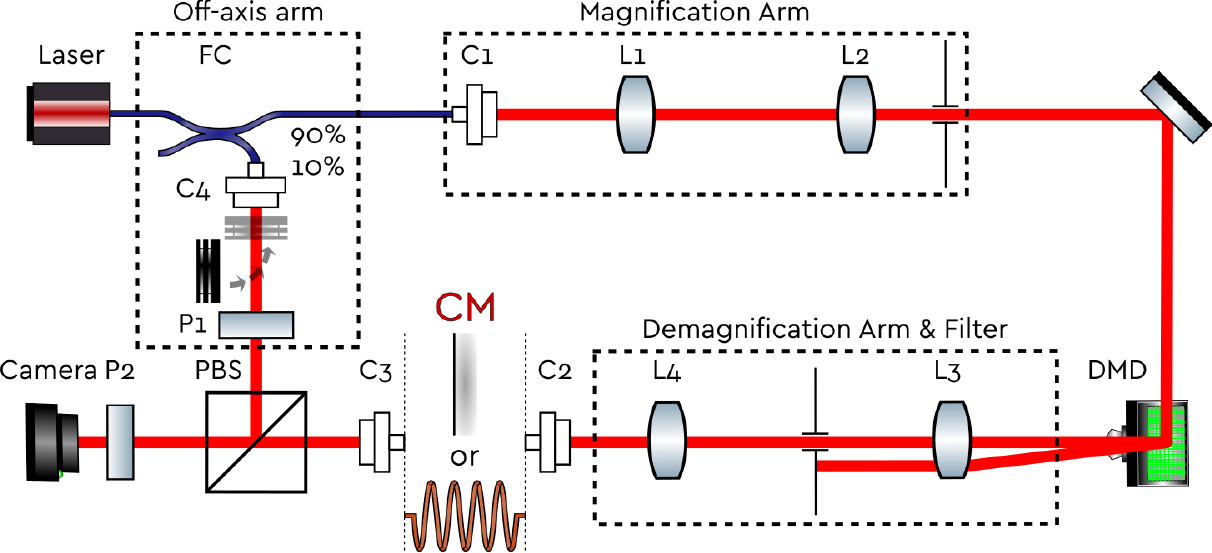}
	\caption{Schematic representation of the setup. 
		A expended laser beam is modulated after reflection off a DMD and injected into a complex medium (CM: ground glass diffuser or multimode fiber).
		One polarization of the outgoing light is recorded by a digital camera.
		A reference arm is used only for the final estimation of the fidelity to measure the complex optical field.
		FC: fiber coupler, Ci (with $i\in[1..4]$): fiber colimator, Li (with $i\in[1..4]$): planoconvex lens, PBS: polarization beam splitter, P1 and P2: polarisers.}
	\label{Setup}
\end{figure}

\textbf{Experimental procedure.}
For both experiments, we modulate the impinging wavefront with a Digital Micromiror Device (DMD). 
We achieve a few-level modulation of the optical amplitude and phase using Lee holograms \cite{Lee_hologram_79} (see Supplementary Methods). 
The modulated beam is then projected onto the complex medium.
We collect the output intensity pattern for a single polarization using a digital camera. 
The setup configuration is depicted in figure~\ref{Setup} and detailed in the Methods. 

First, we estimate the medium's  complex-valued transmission matrix $\mathbf{H}$. 
To remove the need for interferometric measurements, which requires high mechanical stability, thus limiting the versatility of the approach, 
and to reduce the effect of measurement errors, 
we use a phase retrieval algorithm.
It allows recovering the full TM from intensity-only measurements~\cite{dremeau2015reference-less,Metzler_2017}.
We divide the pixel array of the DMD into $N$ groups of pixels or \textit{macropixels}.
We then send a learning set of $7 N$ random binary vectors (entries have zero or unity amplitude) and measure the corresponding output intensity patterns on the camera. The obtained data is fed into a phase retrieval algorithm \cite{Metzler_2017} and the results are further refined using a gradient descent optimization \cite{tensorflow2015-whitepaper} to better account for the non-linearity of the camera response (see Methods and Supplementary Methods). 

Having estimated the complex medium's TM, we go on to divide our input basis into $n$ segments of $N/n$ macropixels, each one corresponding ultimately to one input of the desired operator $\mathbf{G}$ that we want our system to implement.
Identifying an input mask on the DMD that approximates the equality in equation~\ref{eq:A} for a given target matrix $\mathbf{G}$ is an ill-posed problem since we do not have full control over the complex wavefront.
We use a mixed-integer convex solver~\cite{gurobi,cvxpy} to find an approximate solution with discrete amplitude levels that satisfies the experimental restrictions of the binary wavefront modulation of the DMD.
Once the appropriate input projector mask is calculated and displayed on the DMD, the system is ready to act like the desired linear operator. 
In principle, input information is encoded into the optical field values impinging on the different segments of the DMD. 
For the sake of simplicity, we directly encode the information by modulating the mask segments with the DMD, thus limiting ourselves to a few level modulation for the input vector $\mathbf{X}$ (see the detailed procedure in Supplementary Methods).

\section*{Discussion}

We demonstrated the possibility of using  cheap and common media as optical processing units to perform linear operations using wavefront shaping. 
The attractiveness of such approach is linked to its ability to be scaled up for larger operations.
While we report the implementations of $16$ by $16$ operators, much larger operations can be performed provided an increased control over the input field and a reduced noise level.
Noticeably, because of the binary nature of the DMD modulation, equation~\ref{eq:A} was only satisfied approximately and the search for such approximation requires computation efforts.
While averaging over realizations mitigates the error, a single shot operation is usually wanted. 
For an ideal amplitude and phase modulation, equation~\ref{eq:A} can be satisfied exactly with a simple matrix inversion as long as $N \geq n \times m$ ($N \geq n^2$ for a square target matrix) and $M \geq m$ (see Methods). 
A complete independent amplitude and phase modulation can for example be obtained using two liquid crystal spatial light modulators. 
Another approach is, after characterization of the medium, to print phase plates to fix the input mask. 
While limiting the reconfigurability of the system, it reduces the final cost of the processing unit.

For a square target matrix $\mathbf{G}$, the largest size for which one can find a solution for a fixed number $M$ of controlled modes is $n=m=\sqrt{M}$.
Using a multimode fiber as a complex medium, the limiting factor is the number of modes supported by the fiber; for a typical large core step index fiber ($550\ \mu m$ core, $NA = 0.22$), $M\approx 120,000$, corresponding to maximal operator dimensions $n=m\approx350$.
In contrast, in scattering layers such as glass diffusers, the number of degrees of freedom available given by the number of propagating modes is quasi illimited. 
Thus, the number of independently controlled modes is limited by the number of pixels on the modulator, typically of the order of one million. 
Hence, this would allow creating linear operators of size $n=m\approx1000$. 
These operator sizes match the order of magnitude of the size that will be needed for optical neural networks.

It is important to note that our approach, due to intrinsic losses and the absence of gain in the system, can perform any linear operation only up to a constant multiplier. 
In particular, it cannot perform operations with above unity singular values.
Such a restriction can be detrimental to quantum optics applications where losses can modify the optical state of light.
Our apparatus based on DMDs cause more than 50\% of the light to be lost upon modulation.
However, using a phase only spatial light modulator based on deformable mirrors or a phase plate together with a careful injection into a multimode fiber, close to unitary operations can be achieved.

The presented method requires computational efforts during its calibration but once the transmission matrix is retrieved and the projectors are calculated, the operations are performed in a single shot ($O(1)$ operations) on a passive system.
The proposal offers the possibility to implement large optical linear transformations without elaborate fabrication techniques as well as to reconfigure the desired operator without further measurements. 
Moreover, it opens the opportunity to drastically reduce the energy consumption compared to classical electronic components while increasing the computation speed.
These characteristics may enable the presented technique to play a key role in the advent of optical analogue computation and machine learning. 

\section*{Methods}

{\footnotesize
\textbf{Optical configuration.}
We use two different setups for the MMF and scattering medium experiment. 
The MMF one uses a 1 meter long segment of a step index fiber of numerical aperture (NA) 0.22 with a core diameter of 105 $\mu m$ supporting approximatly 4000 guided modes.
The light source is a 1.55$\mu m$ telecom narow-band laser (Teraxion PS-NLL). 
We use a digital micromiror device (Vialux V-9601, 1920 by 1200 pixels, 16 kHz) for the modulation and the output intensity pattern is imaged onto a fast InGaAs camera (Xenics Cheetah 640CL, 640 by 512 pixels). 
The fiber is compressed at four different locations to ensure strong mode couping~\cite{xiong2017principal}. 
The scattering media setup uses a ground glass diffuser (Thorlabs DG20-1500) with a 632.8$nm$ laser source (JDSU 1137/P) modulated by a DMD (Vialux V-9001, 2560 by 1600 pixels, 13 kHz) and a CCD camera (Allied Vision Prosilica GT2300, 2336 by 1752 pixels). 
The MMF setup is depicted in figure~\ref{Setup}, the scattering medium setup is schematically similar. % but without the off-axis interferometric arm. 
The number of controlled segments on the DMD is $N=1568$ for the MMF experiment and $N=2304$ for the scattering media one. 
Both complex media stay highly correlated during the time of the experiment: the output intensity patterns shows correlations greater than 98\% for about two hours (see details in Supplementary Methods).
For the final validation step, we directly measure the output complex field using off-axis holography~\cite{Cuche_99}.
It uses a reference arm originating from a 90/10 fiber splitter at the output of the laser.
The reference beam and the output of the MMF are recombined using a polarizing beam splitter and a polarizer.
The recorded images are post-processed to retrieve the phase information from the interference pattern.

\textbf{Phase retrieval.}
In order to maximize the stability and reliability of the measurements of the transmission matrix \textbf{H}, we choose a full numerical reconstruction of the complex field from intensity measurements using the prVAMP algorithm developed in~\cite{Metzler_2017}. 
The method relies on the recording of a higher number of training samples compared to the number of input elements on the DMD.
Each row of \textbf{H} is then reconstructed independently, allowing parallelization. 
We use $7N$ binary amplitude patterns as training samples to ensure the convergence of the algorithm.
We reconstruct matrices of size $100\times1568$ for the MMF setup and $100\times 2304$ for the scattering medium one. 
The computation is accelerated through parallelization with a graphical processing unit (GPU; NVIDIA GTX 1050 Ti) and is completed in under two minutes. 
We evaluate the quality of the reconstructed transmission matrix using test input wavefronts that have not been used for the learning procedure by calculating the root mean square error (RMSE), giving $RMSE_{mean} \sim 7\%$ ($std(RMSE_{mean}) \sim 6\%$ and $RMSE_{median} < 5\%$) for the MMF setup, and $RMSE_{mean} \sim 11.6\%$ ($std(RMSE_{mean}) \sim 7.3\%$ and $RMSE_{median} < 7.9\%$) for the visible setup. 
As we need a number of outputs channels $m$ lower than the number of output speckle grains measured $M$, we select the ones giving the lowest reconstruction error. 

\textbf{Approximation of the target matrix with projections.}
The spatial light modulator is divided into $n$ segments of size $N/n$ and we select $m$ of the $M$ output measurements.
For illustration purposes, we take $n = m = 4$.
The corresponding projectors $\mathbf{P_{in}}$ and $\mathbf{P_{out}}$ have the following matrix representations:

\begin{equation}
\mathbf{P_{in}} =
\left[
\begin{smallmatrix}
%\smatrix{
p_{1,1} & 0 & 0 & 0\\
\svdots & \svdots & \svdots & \svdots\\
p_{N/4,1} & 0 & 0 & 0\\
0 & p_{1,2} & 0 & 0\\
\svdots & \svdots & \svdots & \svdots\\
0 & p_{N/4,2} & 0 & 0\\
0 & 0 & \sddots & 0\\
0 &0 & 0 & p_{1,m}\\
\svdots & \svdots & \svdots & \svdots\\
0 & 0 & 0 & p_{N/4,m}\\
%}
\end{smallmatrix}
\right]
,\,
\mathbf{P_{out}} =
\left[
\begin{smallmatrix}
\vphantom{\phi_1}1 & 0 & 0 & \shdots & 0\\
\vphantom{\phi_1}0 & 1 & 0 & \shdots & 0\\
\vphantom{\phi_1}0 & 0 & 1 & \sddots & 0\\
\vphantom{\phi_1}0 & 0 & 0 & \sddots & 0\\
\vphantom{\phi_1}0 & 0 & 0 & \sddots & 1\\
\vphantom{\phi_1}0 & 0 & 0 & \shdots & 0\\
\vphantom{\phi_1}\svdots & \svdots & \svdots & \shdots & \svdots\\
\vphantom{\phi_1}0 & 0 & 0 & \shdots & 0
\end{smallmatrix}
\right]
\label{eq:P1_P2}
\end{equation}

\noindent where $p_{k,l}$, $(k,l) \in \llbracket 1,N/4 \rrbracket \otimes \llbracket 1,4 \rrbracket $ represents the modulation on $l^\text{th}$ pixel of the $k^\text{th}$ segments of the modulator.
For the sake of simplicity, the output projection is done here by taking the $4$ first elements of the output basis.
The numerical procedure used to build these projectors is detailed in the Supplementary Methods.

\textbf{Theoretical limit of the scaling.}
We assume the ideal case of $\textbf{H}$ being a Gaussian random matrix measured with negligible noise and the spatial modulation scheme being able to control with a high fidelity the amplitude and the phase of the field.
The Gaussian matrix approximation is known to be valid when one controls and measures a fraction of the total number of modes~\cite{goetschy2013filtering}.
Satisfying the equality of equation~\ref{eq:A} with the projectors as represented by equation~\ref{eq:P1_P2} consists in solving $n$ systems of $m$ linear equations:

%\begin{equation}
\begin{align}
\mathbf{H_{k\perp }} \left[\begin{smallmatrix}p_{1,k}\\ p_{2,k} \\\svdots \\ p_{N/n,k}\end{smallmatrix}\right]
= \mathbf{G_k} \quad \text{for } k \in \llbracket 1,n \rrbracket
\label{eq:gurobi}
\end{align}

\noindent with $\mathbf{H_{k\perp}}  = \mathbf{P_{out}^T}\mathbf{H_k}$ the $m$ by $N/n$ subpart of $\mathbf{H}$ corresponding to the transmission matrix that links the pixels of the $k^\text{th}$ part of the SLM to the $m$ selected outputs on the camera.
$\mathbf{G_k}$ is the $k^\text{th}$ column of the target matrix $\mathbf{G}$.
If a solution exists, it can be written:

\begin{equation}
\left[\begin{smallmatrix}p_{1,k}\\ p_{2,k} \\\svdots \\ p_{N/n,k}\end{smallmatrix}\right]
=
\mathbf{H_{k\perp }^+}\mathbf{G_k}
\label{eq:gurobi_problem}
\end{equation}

\noindent with $\mathbf{H_{k\perp}^+}$ the Moore-Penrose inverse (pseudo-inverse) of $\mathbf{H_{k\perp}}$.
Such a solution exists if $\mathbf{H_{k\perp}}$ has linearly independent rows.
Because $\mathbf{H_{k\perp}}$ is a Gaussian random matrix with  independent identically distributed elements, it can be eigendecomposed and its singular value distribution follows the Marceko-Pastur law~\cite{marcenko1967distribution}.
For large matrices, the probability of having a zero singular value vanishes for $N/n > m$, ensuring that $\mathbf{H_{k\perp}^+}$ exists and is a solution.
If one can control independently the amplitude and the phase of the optical field, the complex mask corresponding to equation~\ref{eq:gurobi_problem} can be implemented. 
We thus find that for a full complex field modulation, it is sufficient to control $M = n \times m$ independent channels of the complex system to be able to correctly simulate any complex target matrix $\mathbf{G}$.

}

\section*{Acknowledgment}
We thank Yaron Bromberg and Arthur Goetschy for fruitful discussions.
We acknowledge the contribution from Christopher Metzler and Laurent Daudet for sharing phase retrieval codes and providing helpful advices.
This work was supported by the French "Agence Nationale pour la Recherche" under Grant No. ANR-16-CE25-0008-01 MOLOTOF and
Labex WIFI (Laboratory of Excellence within the French Program Investments for the Future) under references ANR-10-LABX-24 and ANR-10-IDEX-0001-02 PSL*. P.d.H. acknowledges funding from the French “Ministère de la Défense, Direction Générale de l’Armement”.

\section*{Author Contributions}

The project was initiated by P.d.H., G.L. and S.M.P., conceptualized by P.d.H., S.M.P. and M.W.M., and conducted by M.W.M., P.d.H. and S.M.P. The manuscript was drafted by M.W.M. All authors thoroughly discussed all stages of the project and commented on the manuscript. S.M.P. supervised the project.

\section*{Data Availability}

The datasets generated during the current study are available from the corresponding author upon reasonable request. 
Interface codes for controlling the DMD are made available at~\cite{popoffALP4lib2016} and detailed tutorials on setting up DMD experiments at~\cite{wavefrontshaping_net}.

\clearpage



\section*{Supplementary material}

# Complex Media as Universal Reconfigurable Linear Operators: Supplementary Materials
## Supplementary Methods

M. W. Matthès,[1] P. del Hougne,[1] J. de Rosny,[1] G. Lerosey,[2] and S. M. Popoff[1]

[1]CNRS, ESPCI Paris, PSL Research University,

Institut Langevin, 1 rue Jussieu, 75005 Paris, France.

[2]Greenerwave, ESPCI Paris Incubator PC'up, 6 rue Jean Calvin, 75005 Paris, France

## SETUP STABILITY

Our method relies on projections of the transmission matrix (TM) of a complex medium. It is then crucial that the TM remains constant or highly correlated during the time of the experiment. For optimal projectors $\mathbf{P_{in}}$ and $\mathbf{P_{out}}$, the system mimics the desired operator $\mathbf{G}$ according to equation 2 of the main text. A fluctuation $\mathbf{\Delta H}$ of the TM $\mathbf{H}$ leads to an error $\mathbf{E}$ on the effective operation that reads:

$$\mathbf{E} = \mathbf{P_{out}^T \Delta H \, P_{in}}$$

Highest experimental fidelities are thus obtained by keeping $\mathbf{\Delta H}$ as small as possible, which requires the system not to decorrelate significantly over time. We assess the stability of the medium by calculating the Pearson correlation coefficient between an output image at the beginning of the experiment and output images obtained regularly during the experiment for the same input mask.

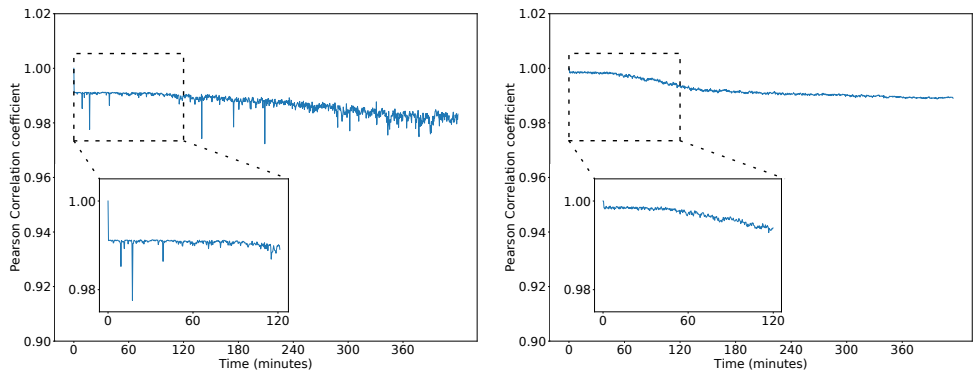

**Figure 1** – Correlation as a function of time, measured for over 6 hours for the MMF setup (left) and the scattering medium setup (right), with a close-up on the first two hours.

In figure 1, we present the evolution over time of the Pearson correlation coefficient for both the Multimode Fiber (MMF) and the scattering medium setups. Both systems stay highly correlated ($> 0.98\%$) over times largely superior to the typical duration of the experiment, which is about 30 minutes.

## DMD MODULATION: THE LEE HOLOGRAM PROCEDURE

Digital Micromirror Devices (DMDs) allow binary amplitude modulation (*on/off*) of an input beam by tilting numerous small mirrors, used as pixels, into two possible orientations, directing light either in or out of the optical setup. While these devices allow fast modulation ($> 10kHz$)

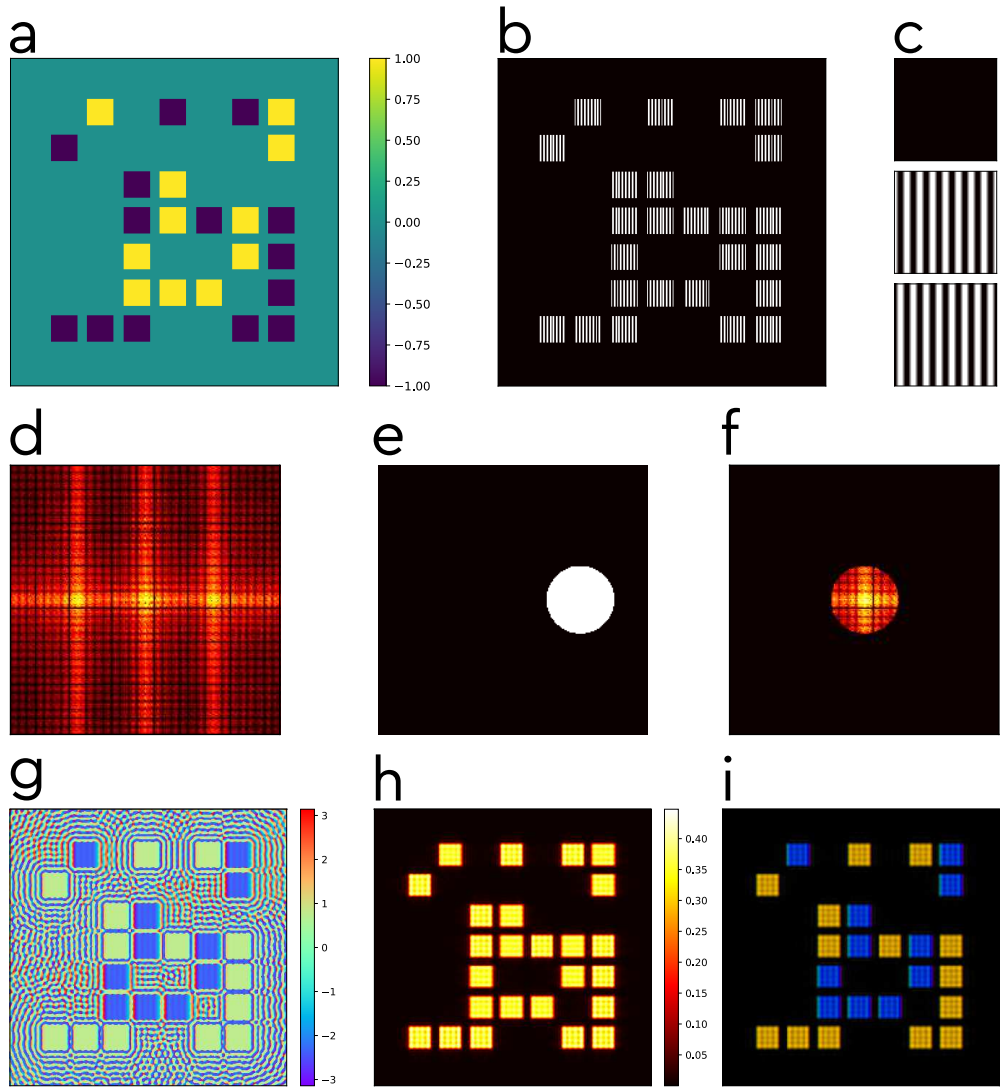

**Figure 2** – Numerical simulations of the Lee hologram procedure: **a**: a desired three level optical field modulation. **b**: Corresponding image displayed on the DMD. **c**: Close up of the three macropixel patterns corresponding to a field value of 0, −1 and 1 respectively. **d**: Fourier transform of **b** (log scale), mimicking the effect of a lens. **e**: Filtering mask used to keep the first order replica, simulating the experimental spatial frequency selection by an iris. **f**: Filtered and centered first order replica (log scale). **g** and **h**: Phase and amplitude of the inverse Fourier transform of **f**. **i**: HSV representation of the final field.

compared to liquid crystal modulators ($\leqslant 100 Hz$), they are limited by the lack of modulation depth and by their inherent incapacity to modulate the phase of the optical field. Moreover, particular attention should be paid to the diffraction effects as a DMD acts as a blazed grating, imposing to carefully choose the pixel size and the angle of the input beam for a given wavelength [1]. Simultaneous amplitude and phase modulation can be obtained by using the Lee Hologram modulation approach [2]. It consists in displaying modulated patterns on the DMD and filtering the spatial

frequencies in the Fourier plane of a lens. This allows creating a phase modulation in the image plane of the optical system. Moreover, any unwanted direct reflection is filtered out by the system and does not contribute to the output signal, thus increasing the contrast of the amplitude modulation. In our experiment, we used a depth of two for the phase modulation. This allows creating two phase levels $\phi \in \{0, \pi\}$, giving access to three amplitude values $V \in \{-1, 0, 1\}$. We present in figure 2 the different steps of the Lee Hologram method using square groups of pixels on the DMD.

## PHASE RETRIEVAL TECHNIQUES

As evoked in the main text, the advantage of using numerical phase retrieval techniques is to not rely on interferometry, which is a source of instability and reconstruction noise. For the characterization of the transmission matrix, only the optical intensity is measured thanks to a digital camera. In figure 3 we show two typical random input patterns used for the phase retrieval sequences, two output speckles recorded on the camera and their corresponding vectorization onto the output basis which consists in hexagonal groups of pixels.

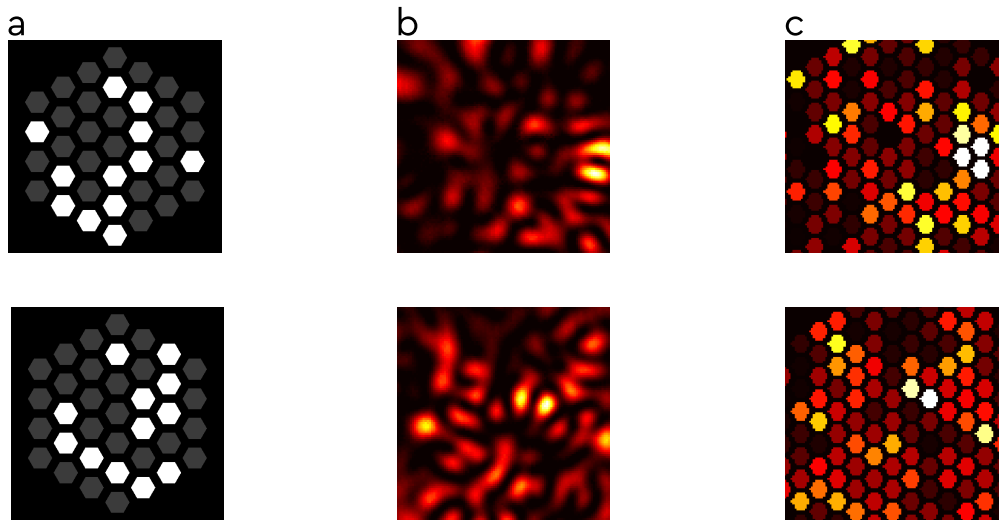

**Figure 3** – **a**: Sample of two random input patterns composed of hexagonal macropixels used in the TM learning sequence. **b**: Collected intensity speckles on the camera. **c**: Projections onto the output basis of the speckles shown in **b**.

$7 \times N$ pairs of input-output vectors are used for the phase retrieval procedure, where $N$ is the size of the input basis. This number is empirically chosen as a tradeoff between measurement speed and reconstruction efficiency [3]. To reconstruct the phase from intensity information, we use algorithms based on Approximate Message Passing (AMP) which is a class of heuristic algorithms originally

designed to solve compressed sensing problems [4]. We use the *prVAMP* implementation [3] which demonstrated improved computational times compared to the previous *prSAMP* approach [5] while keeping the reconstruction efficiency. We run the parallel version of code used in [3] and perform the computation on a graphical processing unit (NVIDIA GTX 1050 Ti) to divide computational time by a factor ranging from 20 to 40 for a typical matrix size of $100 \times 1568$ compared to the same calculations on the central processing unit.

The sole usage of this algorithm was enough to reconstruct the transmission matrix with a high fidelity for the MMF setup. However, we observed more significant reconstruction errors for the scattering medium experiment. We estimate the source of error to be the linearity of the pixel response of the CCD, and more particularly the existence of an offset value, i.e. the value returned by a pixel without any input light, that is not constant over the whole CCD array. To overcome this issue, we added a complementary step after the phase retrieval attempt to *learn* this offset. We implemented a mini-batch gradient descent [6] to refine transmission matrix $\mathbf{H}$ and find the best offset-correction vector $\mathbf{W}$. The problem can be writen as:

$$(\mathbf{H}, \mathbf{W}) = \underset{\mathbf{H}, \mathbf{W}}{\operatorname{argmin}} \, \overline{F(\mathbf{H}, \mathbf{W})}$$

$$\text{with } \overline{F(\mathbf{HW})} = \frac{1}{7N} \sum_{i=1}^{7N} \| \mathbf{HX_i} + \mathbf{W} - \mathbf{Y_i} \|_2$$

with $(\mathbf{X_i}, \mathbf{Y_i})_{i \in (1..7N)}$ the pairs of input and output vectors obtained in the training sequence. We initiated the gradient descent with the approximation of the TM $\mathbf{H}$ previously obtained with *prVAMP*. Refining the TM as well as taking into account the obtained offset vector allowed a reduction of the reconstruction error of the TM, going from $RMSE_{mean} = 15.7\% \pm 7.9\%$ and $RMSE_{median} = 14.4\%$, to $RMSE_{mean} = 11.6\% \pm 7.3\%$ $RMSE_{median} = 7.9\%$.

## ENCODING INPUT VECTORS ONTO THE PROJECTORS

Our setup allows the computation of a linear operation applied to an input vector $\mathbf{X}$ of size $n$ encoded on $n$ incoming light beams. This approach is compatible with an integration into a larger optical processing unit. In the left part of figure 4, we present this approach: the DMD divided into $n = 4$ quadrants that each receives a single light beam corresponding to an element of the input vector. For the sake of simplicity, rather than using multiple light beams, we encode input vectors by modulating globally each quadrant of the DMD as shown in the right part of figure 4.

This allows using a plane wave as the incident beam. We can express the output optical field $\mathbf{Y}$ as:

$$\mathbf{Y} = \mathbf{GX}$$

$$= \mathbf{P_{out}^T H P_{in} X}$$

$$= \mathbf{P_{out}^T H P_{in} \, diag(X) E}$$

$$= \mathbf{P_{out}^T H \widetilde{P}_{in}(X) \, E}$$

where $\mathbf{diag(X)}$ is the matrix of diagonal elements equal to the elements of $\mathbf{X}$ and the off-diagonal equal to 0, $\mathbf{E} = (1)_n$ a n-size vector of elements all equal to 1, representing the plane wave, and $\mathbf{\widetilde{P}_{in}(X)} = \mathbf{P_{in} diag(X)}$ is the actual mask displayed on the DMD for each individual input vector $\mathbf{X}$.

Using this method, we can express any input vector $\mathbf{X}$ as long as its values are within the modulation depth of our SLM. A schematic comparison of the two processes is presented in figure 4

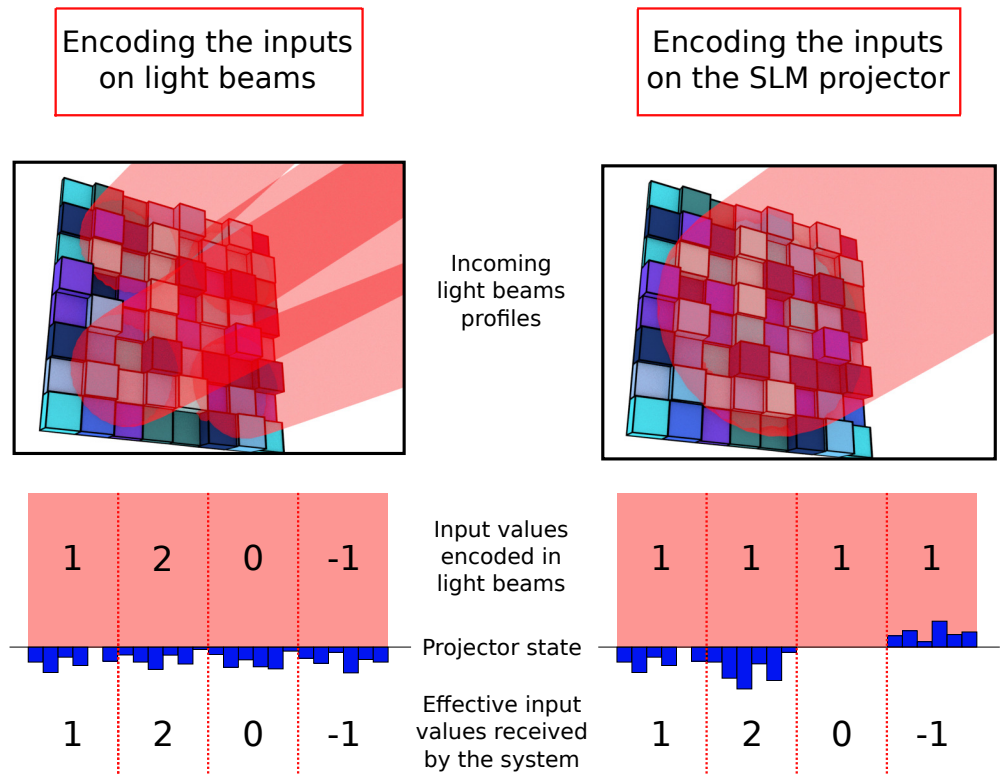

**Figure 4** – Comparison of the two presented methods to encode input vectors in the case of a $4 \times 4$ operator. **Left**: The input vector is brought by 4 light beams that carry the signal information. The input projector obtained in the optimization procedure is then applied to the whole incoming field. **Right**: A single light beam illuminates the SLM. The input vector is built by modulating each sub-part of the projector according to the respective input vector value. The actual mask thus changes for each different input vector.

# PHASE ACQUISITION

While the calibration of the system does not require to measure the optical phase, as relying on numerical phase retrieval, we do measure the phase to estimate the quality of the performed operation compared to the target one. We used a reference arm to perform off-axis holography [7]. The reference is taken from a 90/10 fiber coupler placed directly after the laser source. It is brought to the second entrance of the polarization beam splitter though a polarizer (P1 in the main text, used to tune the intensity of the beam) and goes to the camera with a slight angle with respect to the optical axis. After passing through polarizer P2, the reference and the signal interfere on the camera resulting in a speckle pattern modulated with high spatial frequency fringes. The collected images are then treated numerically. The phase reconstruction consists in selecting the spatial frequencies corresponding to the complex image shifted by the spatial frequency of the fringes [7]. We first perform a discrete Fourier transform of the image. We then apply a filter to select the spatial frequencies corresponding to the first-order replica of the image. The central frequency is obtained by calculating the center of mass of the first order replica. Finally, we shift the filtered images back to the center of the spatial frequency coordinate system and perform an inverse discrete Fourier transform. The complex resulting image has the phase information of the output speckle pattern. We project the obtained complex field onto the output basis (hexagonal groups of pixels) with the same procedure as for the intensity speckles: we calculate the complex average value of the field in each macropixel of the basis.

At this point, it is important to mention that each row of the retrieved TM is obtained up to a global phase and up to a phase conjugation operation. Indeed, the resulting output intensity does not change if a row is conjugated or phase shifted. It is therefore necessary to remove phase uncertainties of the TM before comparing phase measurements to predictions. To do so, we send a sample of test vectors and measure the complex output field using the previously described method. For each output point, we compare the experimental phase to the phase predicted by the TM, the difference between the two phases should be constant within the experimental error range. We evaluate the phase variations by taking the standard deviation of the complex exponential of the phase difference. When this quantity gets higher that 0.5 (corresponding to conjugation mismatch), we phase-conjugate the corresponding row of the TM and recalculate the phase differences. They almost always become constant after the conjugation: using this method we correct the TM to obtain its real phase values. Modifying the TM in this fashion does not change the previously calculated projectors as the rows are modified globally, therefore the results only present a global

phase shift, that we numerically remove before comparing to the predictions obtained with the target operator.

## CALCULATION OF OPTIMAL PROJECTORS

The projectors $\mathbf{P_{in}}$ and $\mathbf{P_{out}}$ we seek are solutions to the equations:

$$\mathbf{P_{out}^T H_k P_{in}} = \mathbf{G_k}, \, k = 1..n \tag{1}$$

Where $\mathbf{G_k}$ is the k$^{\text{th}}$ column of the desired operator, and $\mathbf{H_k}$ is k$^{\text{th}}$ sub-part of the TM, composed of columns with indices going from $k * N/n$ to $(k+1) * N/n - 1$. Our modulation scheme imposes some constraints on the possible values of the superpixels. It imposes the components of $\mathbf{P_{in}}$ to be $-1$, $0$ or $1$. Therefore, $\mathbf{P_{in}}$ cannot be simply obtained by inverting equation 1 as it would result in arbitrary complex values. However, we can find an approximate solution satisfying our constraints, provided that we have enough degrees of freedom $N$. Numerically, we try to find the projectors $\mathbf{P_{in}}$ and $\mathbf{P_{out}}$ that satisfy :

$$(\mathbf{P_{in}}, \mathbf{P_{out}}) = \operatorname*{argmin}_{\mathbf{P_1}, \mathbf{P_2}} \left\| \mathbf{P_2^T H P_1 - G} \right\|_2$$

The output projector $\mathbf{P_{out}}$ selects the output points that will be used as the outputs of the operation. It amounts to selecting rows of the TM. We simply keep the rows that show the lowest reconstruction error. Having selected an output projector, we want to find the input one that satisfies:

$$(\mathbf{P_{in}}) = \operatorname*{argmin}_{\mathbf{P}} \left\| \mathbf{P_{out}^T H P - G} \right\|_2$$

which is a convex problem. To numerically solve it, we use CVXPY, a convex optimization framework for Python [8] which allows us to express in simple terms the problem and to add the discrete value constraint to the elements of $\mathbf{P_{in}}$. We use a multiplicative constant $\gamma$ to tune the average value of the expected signals in order to fit the exposure time and the dynamic range of the camera. The final problem consists in solving $n$ sub-problems:

$$\left\| \mathbf{P_{out}^T H P_k}/\gamma - \mathbf{G_k} \right\|_2$$

We use Gurobi [9] as the backend solver for CVXPY as it allows resolution of mixed integer problems, i.e. finding a solution with discrete values. We engineered the value of $\gamma$ to have the outputs of the operator close the mean intensity of the speckle grains. This choice empirically gives the best experimental results.

We finally evaluate the quality of the optimization by calculating the fidelity $F'_c = Tr(|\tilde{\mathbf{G}}'.\mathbf{G}^\dagger|^2)*$ $n$. $\tilde{\mathbf{G}}'$ is the operator obtained after numerically applying the projectors $\mathbf{P^T_{out}}$ and $\mathbf{P_{in}}$ to the measured transmission matrix $\mathbf{H}$. We always obtained $F'_c > 0.99$, demonstrating the efficiency of the numerical optimization.

## SUPPLEMENTARY RESULTS

**Multimode fiber results.** This section is devoted to presenting the results for the various experiments we conducted with the MMF setup. It includes results for the two operators we presented in the main text, of sizes $n = 8$ and $n = 16$, with our without averaging, and a special operator that we designed to further prove the tunability of our system in figure 5.

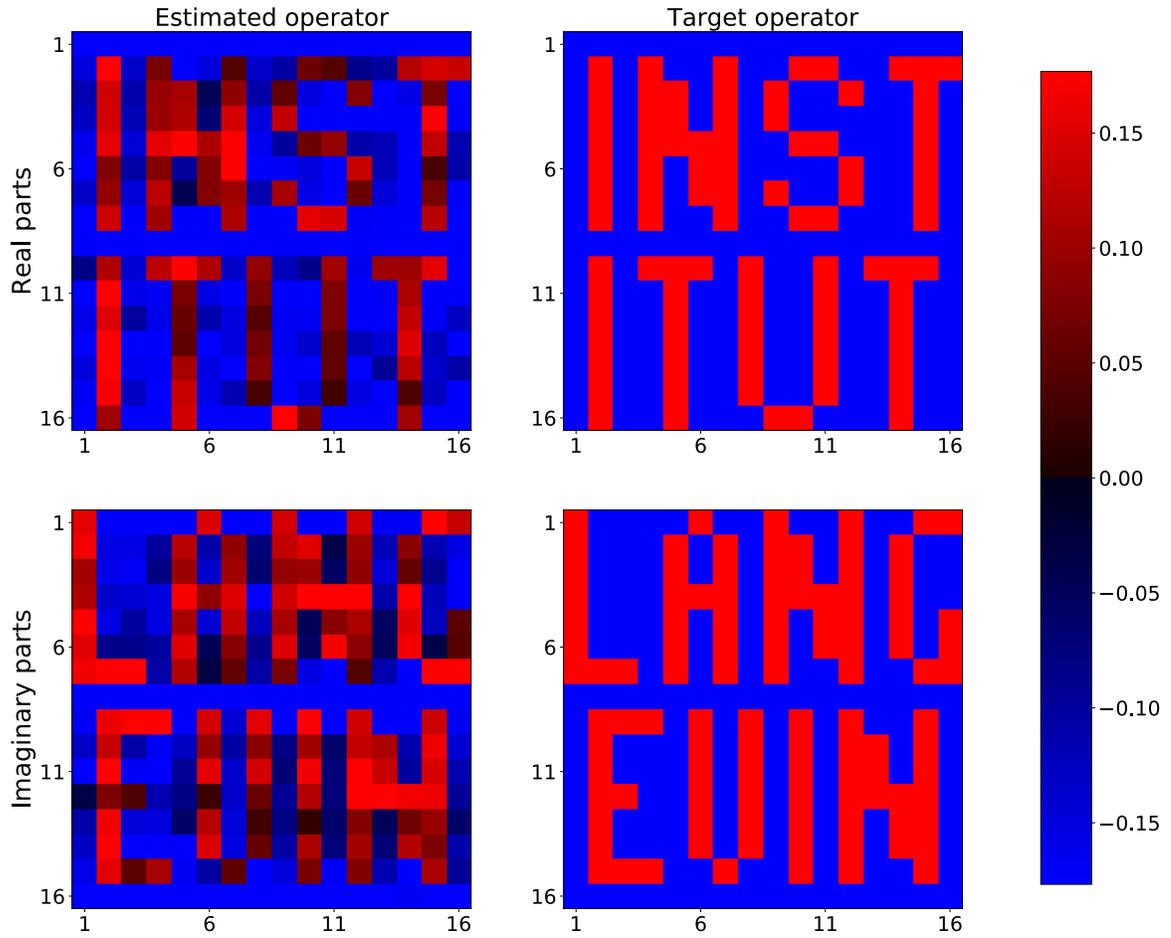

**Figure 5** – Comparison between estimated (left) and target (right) operator, for a designed operator **G**.

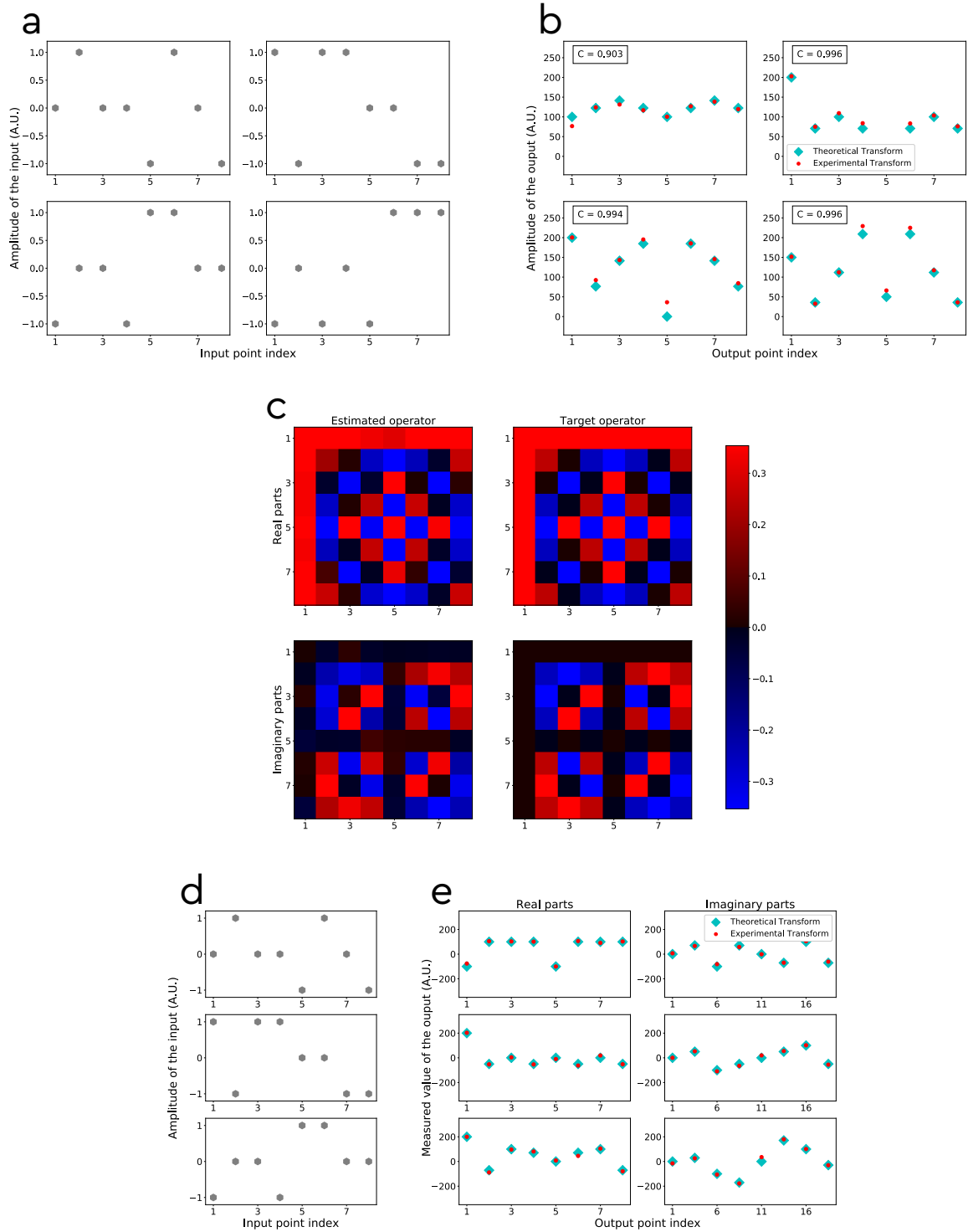

**Figure 6** – Results for **G** = DFT$_8$ obtained in single shots experiments. **a.** 4 different input vectors and **b** the corresponding theoretical and experimental amplitude measurements of the output vectors. **c.** Comparison between estimated (left) and target (right) operator **G**. **d.** 3 different input vectors and **b.** the corresponding real and imaginary parts measurements of the output vectors.

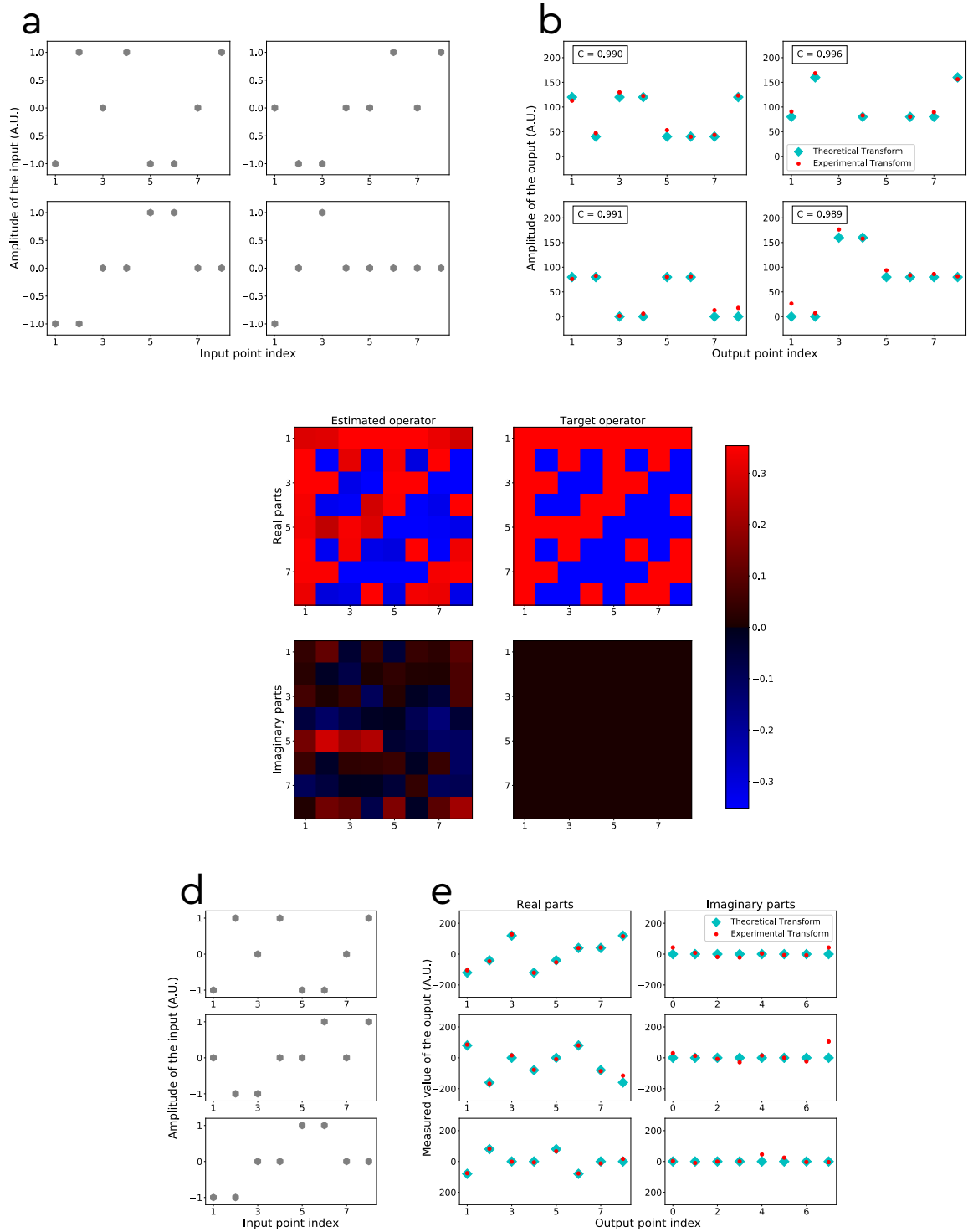

**Figure 7** – Results for **G** = Hadamard$_8$ obtained in single shots experiments. **a.** 4 different input vectors and **b.** the corresponding theoretical and experimental amplitude measurements of the output vectors. **c.** Comparison between estimated (left) and target (right) operator **G**. **d.** 3 different input vectors and **b.** the corresponding real and imaginary parts measurements of the output vectors.

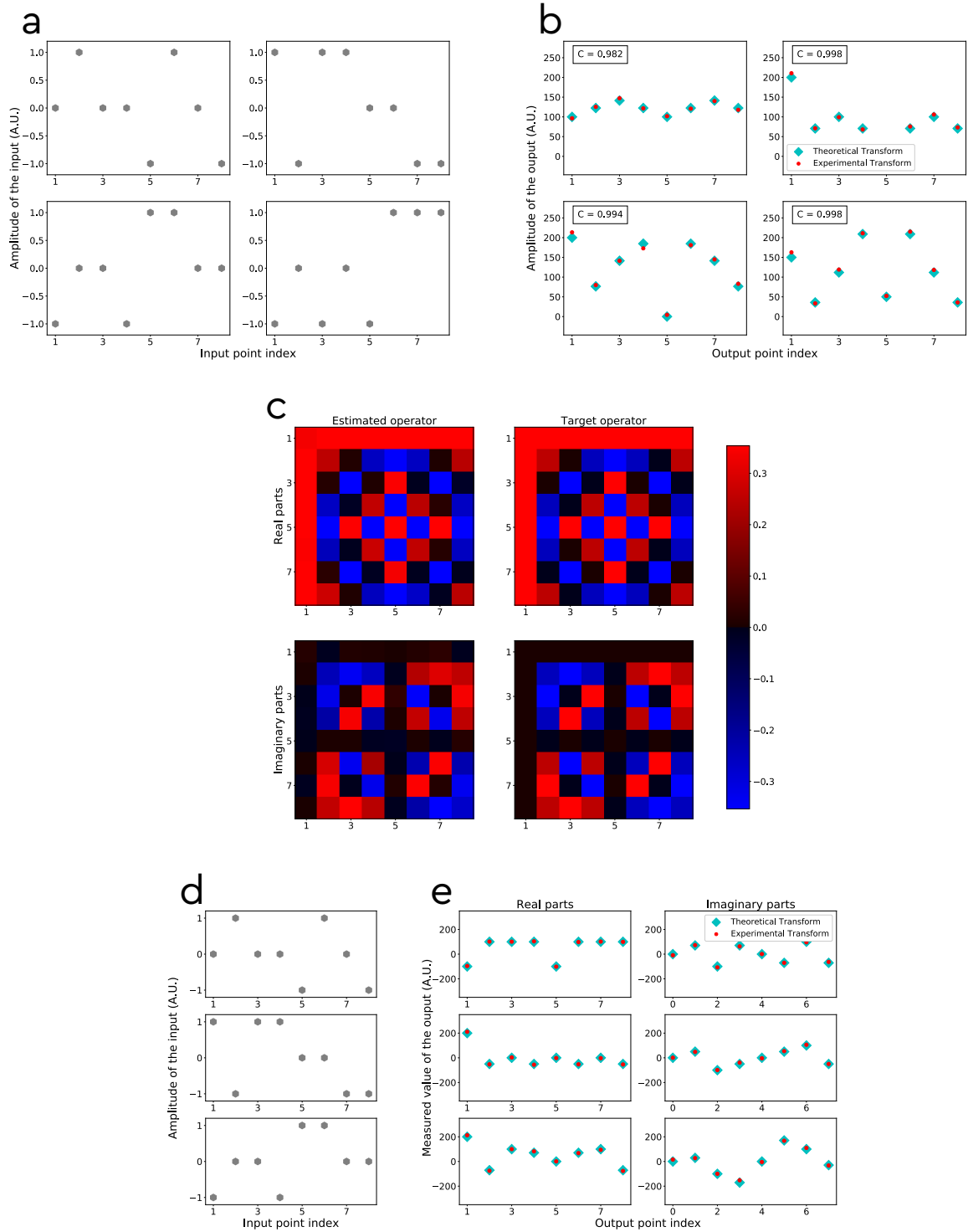

**Figure 8** – Results for **G** = DFT$_8$ obtained after averaging over 5 experiments. **a.** 4 different input vectors and **b.** the corresponding theoretical and experimental amplitude measurements of the output vectors. **c.** Comparison between estimated (left) and target (right) operator **G**. **d.** 3 different input vectors and **b.** the corresponding real and imaginary parts measurements of the output vectors.

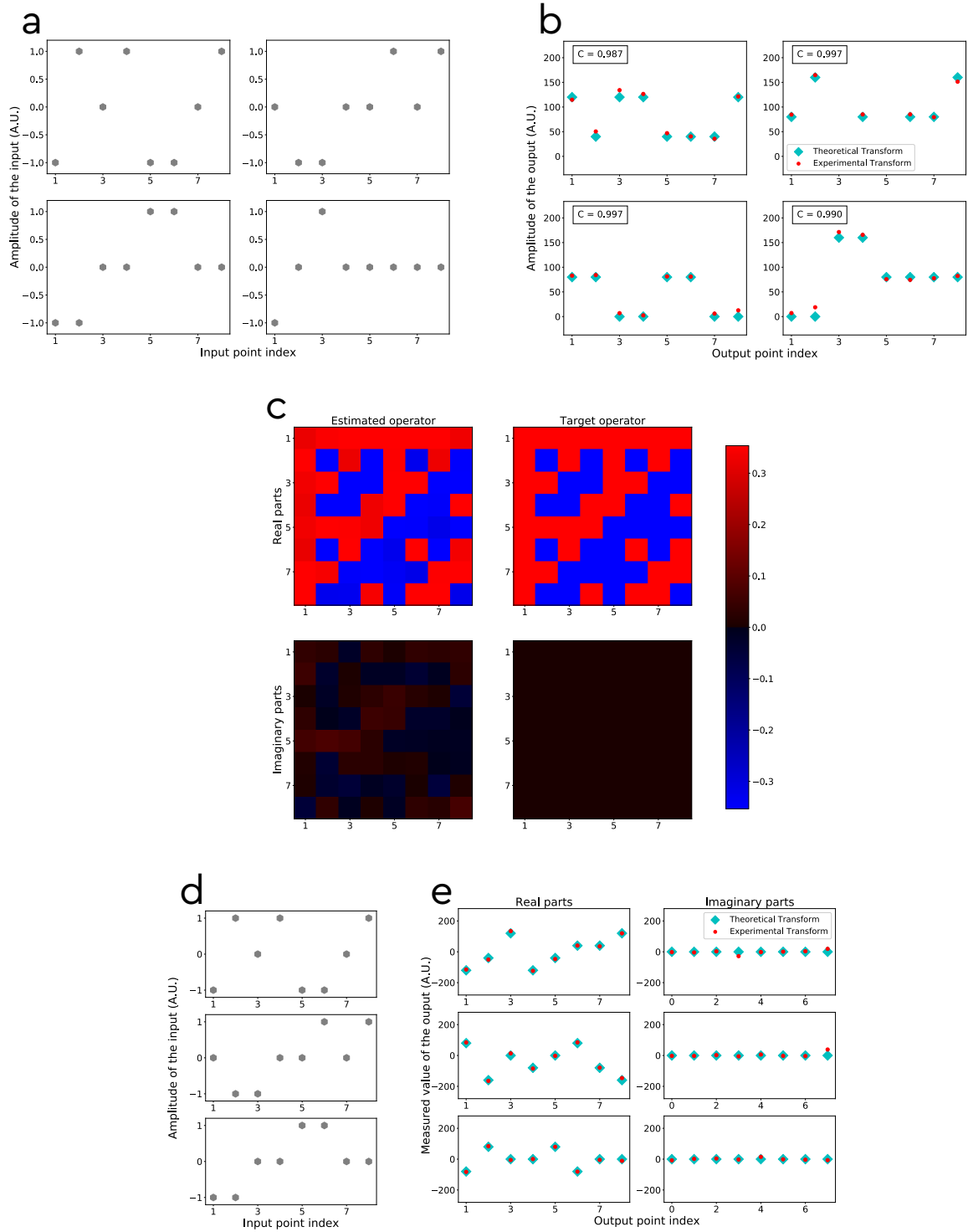

**Figure 9** – Results for **G** = Hadamard$_8$ obtained after averaging over 5 experiments. **a** 4 different input vectors and **b** the corresponding theoretical and experimental amplitude measurements of the output vectors. **c** Comparison between estimated (left) and target (right) operator **G**. **d** 3 different input vectors and **b** the corresponding real and imaginary parts measurements of the output vectors.

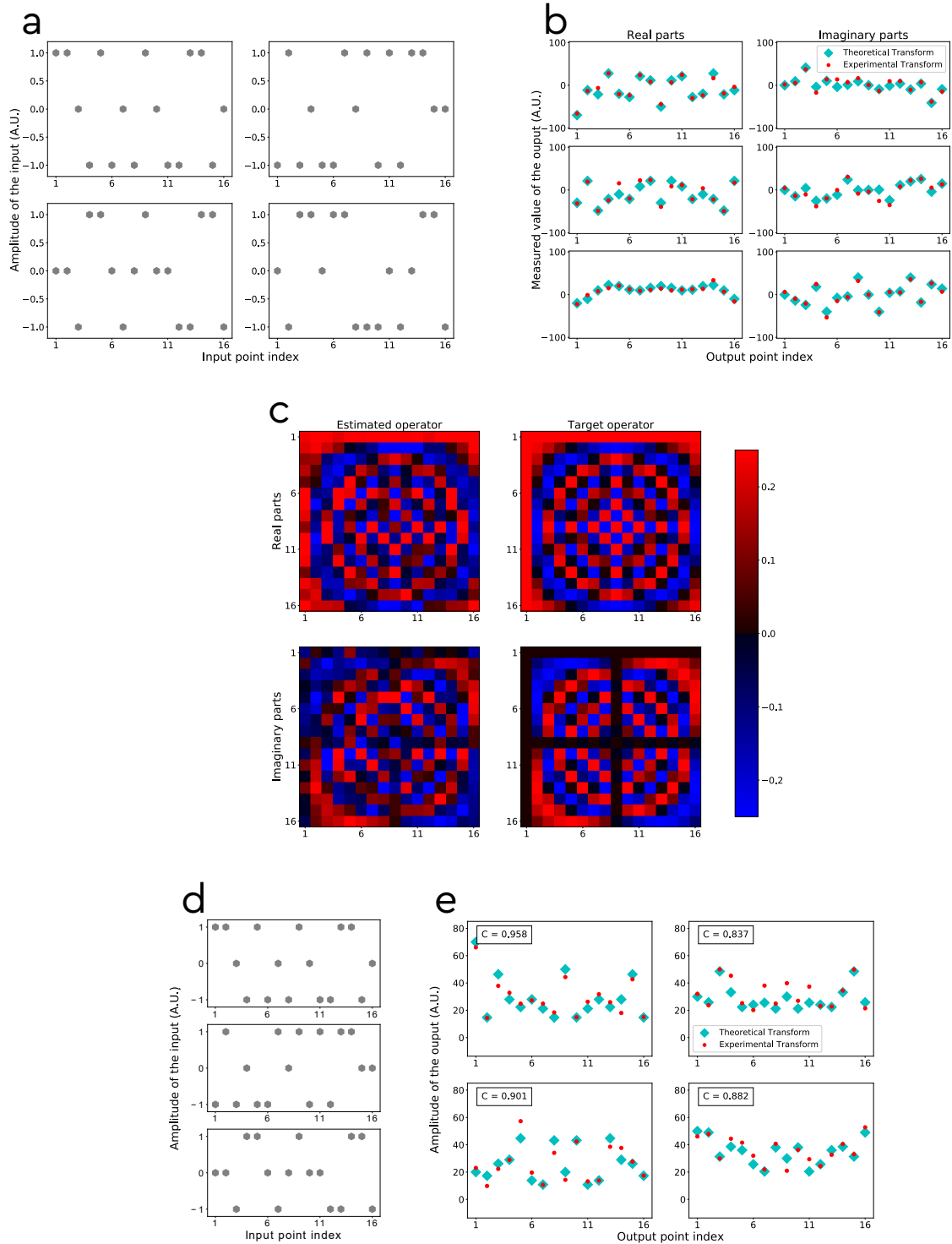

**Figure 10** – Results for **G** = DFT$_{16}$ obtained in single shots experiments. **a.** 4 different input vectors and **b.** the corresponding theoretical and experimental amplitude measurements of the output vectors. **c.** Comparison between estimated (left) and target (right) operator **G**. **d.** 3 different input vectors and **b.** the corresponding real and imaginary parts measurements of the output vectors.

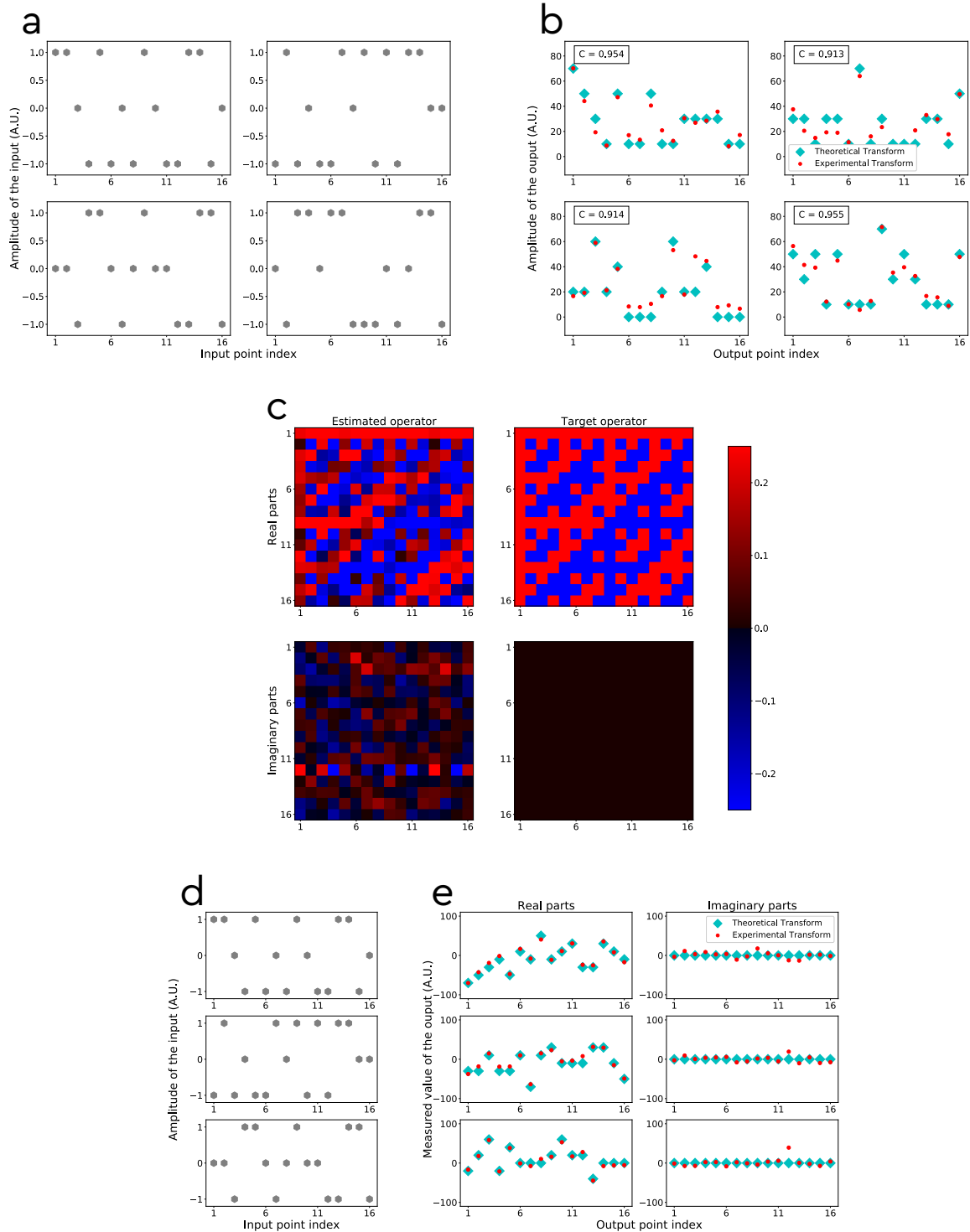

**Figure 11** – Results for $\mathbf{G} = \text{Hadamard}_{16}$ obtained in single shots experiments. **a.** 4 different input vectors and **b.** the corresponding theoretical and experimental amplitude measurements of the output vectors. **c.** Comparison between estimated (left) and target (right) operator $\mathbf{G}$. **d.** 3 different input vectors and **b.** the corresponding real and imaginary parts measurements of the output vectors.

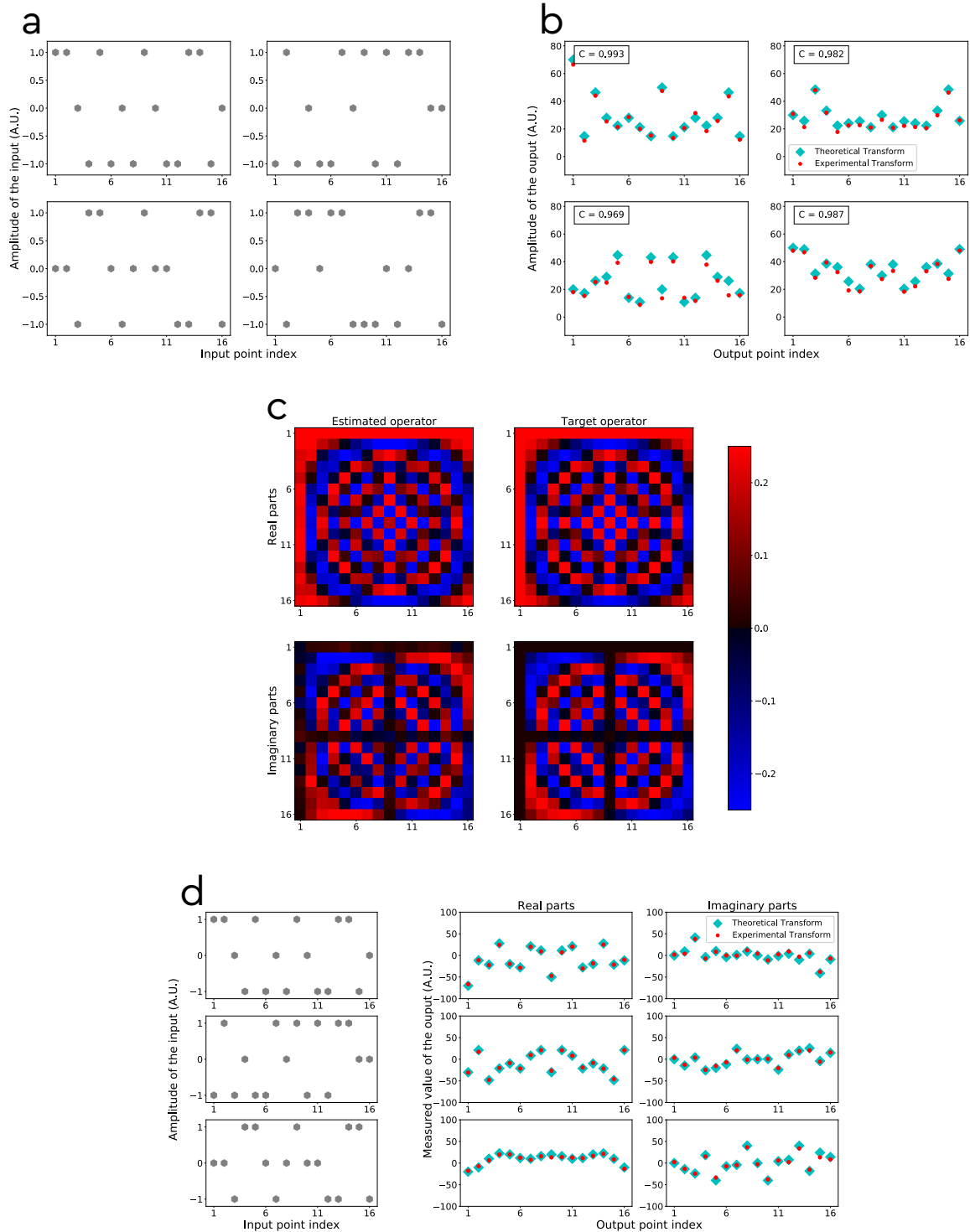

**Figure 12** – Results for **G** = DFT$_{16}$ obtained after averaging over 10 experiments. **a** 4 different input vectors and **b** the corresponding theoretical and experimental amplitude measurements of the output vectors. **c** Comparison between estimated (left) and target (right) operator **G**. **d** 3 different input vectors and **b** the corresponding real and imaginary parts measurements of the output vectors.

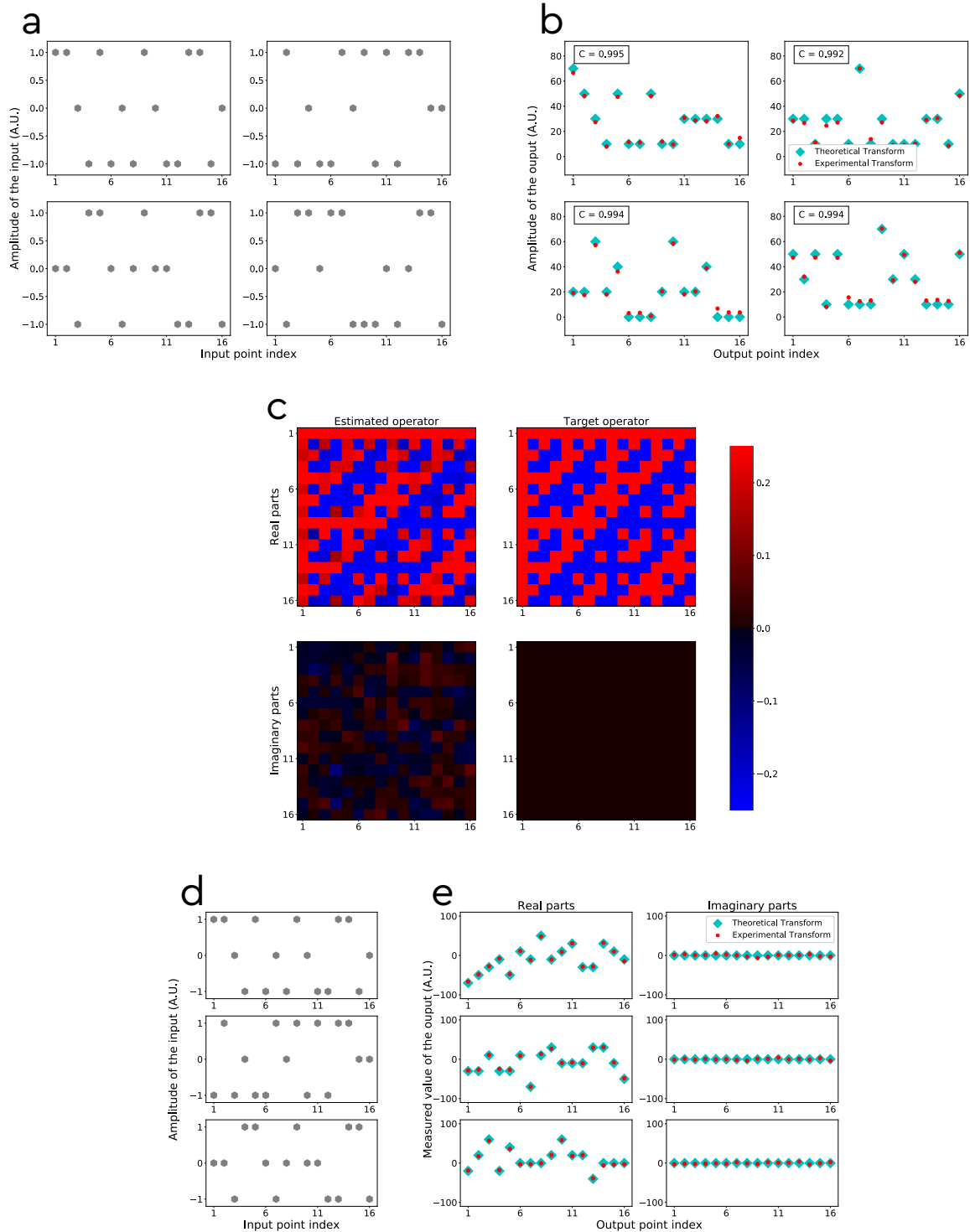

**Figure 13** – Results for **G** = Hadamard$_{16}$ obtained after averaging over 10 experiments. **a.** 4 different input vectors and **.b** the corresponding theoretical and experimental amplitude measurements of the output vectors. **c.** Comparison between estimated (left) and target (right) operator **G**. **d.** 3 different input vectors and **b.** the corresponding real and imaginary parts measurements of the output vectors.

**Scattering medium results.** We present single-shot experimental results with the DFT$_8$ operator.

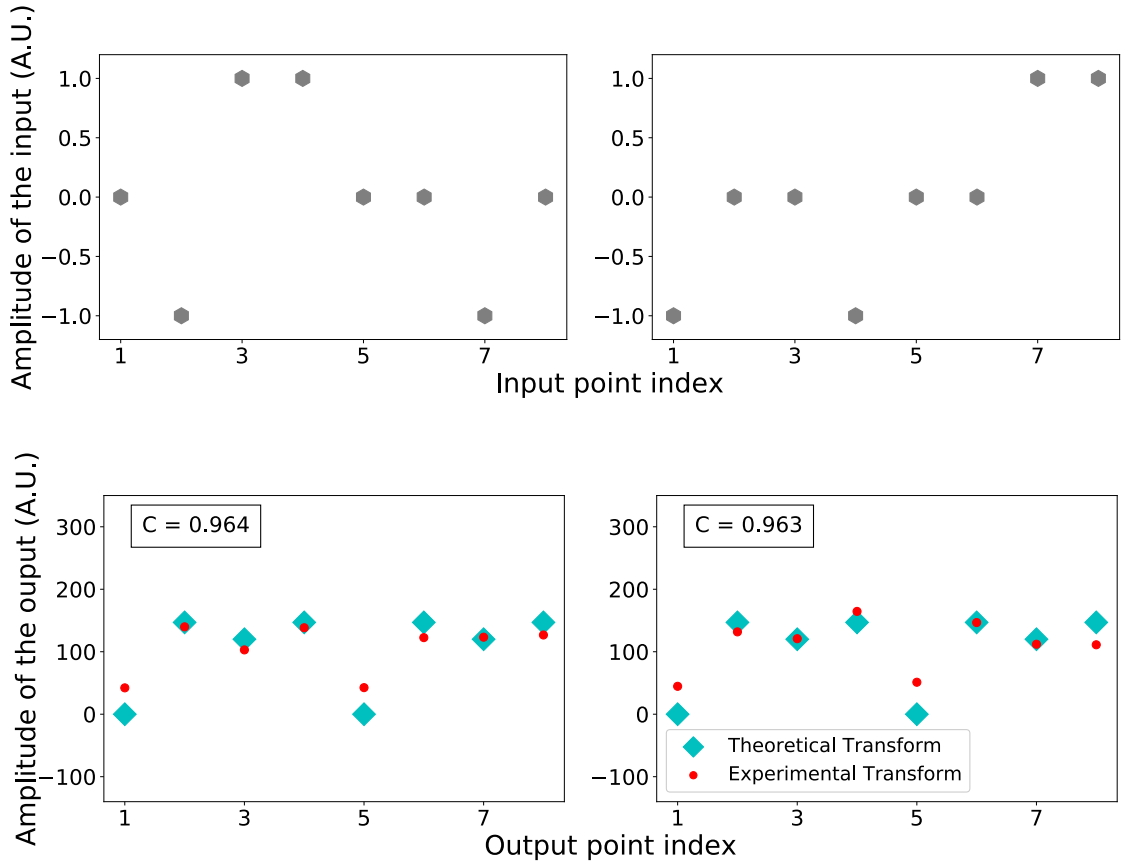

**Figure 14** – Input (top) and corresponding output (bottom) vectors of a single shot discrete Fourier transform of size 8.

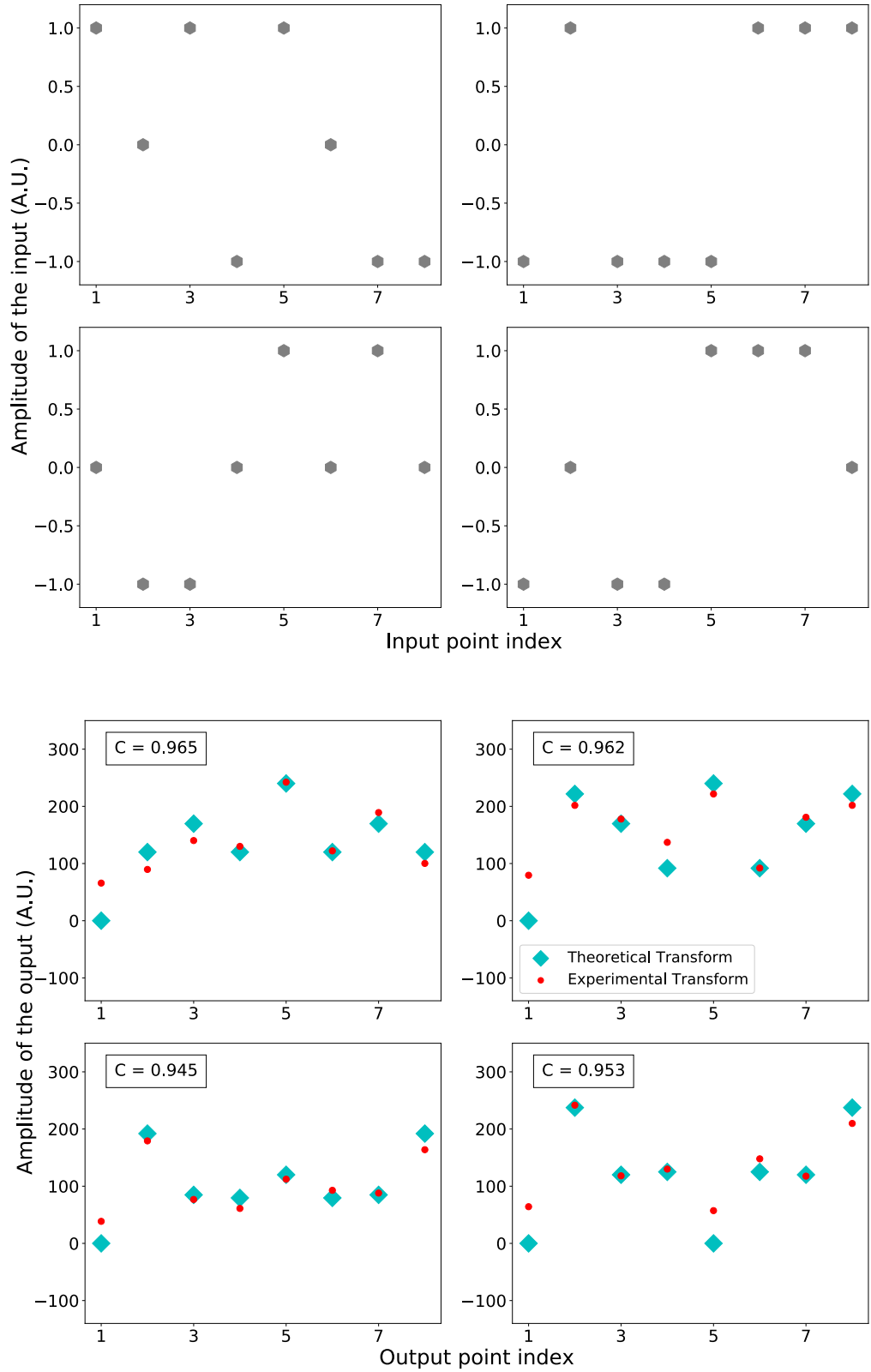

**Figure 15** – Input (top) and corresponding output (bottom) vectors of a single shot discrete Fourier transform of size 8.



\end{document}